\documentstyle [bezier,12pt] {article}
\title{Rewriting Preserving Recognizability of Finite 
Tree Languages}
\author{S\'andor V\'agv\"olgyi\thanks{E-mail: vagvolgy@inf.u-szeged.hu.}
\\{\sl Department of Foundations of Computer Science}
\\{\sl University of Szeged}
\\{\sl Szeged, \'Arp\'ad t\'er 2}
\\{\sl H-6720 Hungary}}
\date{}
\let\s=\sigma    \let\c=\circ     \let\b=
\Box
\let\S=\Sigma   \let\D=\Delta

\def\ts{T_\Sigma}

\def\les#1#2#3{#1\leq #2\leq #3}

\def\seq#1#2#3{#1_{#2},\ldots,#1_{#3}}

\def\red#1{\mathop\rightarrow_{#1}}
\def\tred#1{\mathop\rightarrow_{#1}^{*}}

\def\tthue#1{\mathop\leftrightarrow_{#1}^{*}}

\def\pr{{\bf Proof.\ }}

\newtheorem{tet}{Theorem}[section]
\newtheorem{cla}[tet]{Claim}
\newtheorem{exa}[tet]{Example}
\newtheorem{lem}[tet]{Lemma}
\newtheorem{obs}[tet]{Observation}
\newtheorem{sta}[tet]{Statement}
\newtheorem{prop}[tet]{Proposition}

\newtheorem{df}[tet]{Definition}

\newtheorem{cons}[tet]{Consequence}

\begin{document}
\date{}
\maketitle

\begin{abstract} 
We show that left-linear generalized semi-monadic TRSs 
 effectively preserving recognizability of finite tree languages (are EPRF-TRSs). 
We show that  reachability, joinability, and local confluence are    decidable for 
  EPRF-TRSs. 
\end{abstract}


\section{Introduction}
The notion of preservation of recognizability through rewriting is 
a widely studied concept in term rewriting, see \cite{coq}-\cite{ful}, \cite{fjsv}, \cite{ gill}-\cite{vag}. 
Let $\S$ be a ranked alphabet, let
$R$ be a term rewrite system (TRS) over $\S$, and let $L$ be a tree language over $\S$. 
Then $R^*_\S(L)$ denotes the  set of descendants of trees in
$L$. 
A TRS $R$ over $\S$ preserves $\S$-recognizability (is a P$\S$R-TRS), if
for each recognizable tree language $L$ over $\S$,
 $R^*_\S(L)$ is recognizable.
A TRS $R$ over $\S$ preserves $\S$-recognizability  of finite tree languages (is a P$\S$RF-TRS), if
for each finite tree language $L$ over $\S$,
 $R^*_\S(L)$ is recognizable.

Let $R$ be a TRS over $\S$. Then its signature, $sign(R)\subseteq \S$ is the
ranked alphabet consisting of all symbols appearing in the rules of $R$.
A TRS $R$ over $sign(R)$ preserves recognizability (is a PR-TRS), if 
for each ranked alphabet $\S$ with $sign(R)\subseteq \S$, 
$R$, as a TRS over $\Sigma$, preserves $\S$-recognizability. 
A TRS $R$ over $sign(R)$ preserves recognizability   of finite tree languages (is a PRF-TRS), if 
for each ranked alphabet $\S$ with $sign(R)\subseteq \S$, 
$R$, as a TRS over $\Sigma$, preserves $\S$-recognizability of finite tree languages.

A TRS $R$ over $\S$  effectively preserves $\S$-recognizability (is an EP$\S$R-TRS), if for a given 
 a bottom-up tree automaton (bta)
$\cal B$ over $\S$, we can effectively construct a bta 
$\cal C$ over $\S$ such that $L({\cal C})=R^*_\S(L({\cal B}))$.
A TRS $R$ over $\S$ effectively preserves $\S$-recognizability  of finite tree languages (is an EP$\S$RF-TRS), 
if for a given finite tree language $L$ over $\S$, we can effectively construct a bta 
$\cal C$ over $\S$ such that $L({\cal C})=R^*_\S(L)$.   
A TRS $R$ over $sign(R)$ effectively preserves recognizability of finite tree languages (is an EPRF-TRS),  
if for a given ranked alphabet 
$\S$ with $sign(R)\subseteq \S$ and a given 
finite tree language $L$ over $\S$, we can effectively construct a bta 
$\cal C$ over $\S$ such that $L({\cal C})=R^*_\S(L({\cal B}))$.

Gyenizse and V\'agv\"olgyi \cite{gyevag} presented a linear TRS $R$ over $sign(R)$
such that $R$ is  an EP$sign(R)$R-TRS and 
 $R$ is not a PR-TRS.
 V\'agv\"olgyi \cite{vag}  showed that 
it is not decidable for a murg TRS $R$ over $\S$ whether 
$R$  is a   P$\S$RF-TRS. 
Let $R$ be a rewrite system over $sign(R)$, and let 
$\S=\{\, f, \sharp \,\}\cup sign(R)$, where 
 $f\in \S_2-sign(R)$ and $\sharp\in \S_0-sign(R)$. 
Gyenizse and V\'agv\"olgyi \cite{gyevag} showed that
 $R$ is an EP$\S$R-TRS  if and only if 
$R$  is an EPR-TRS.
Gyenizse and V\'agv\"olgyi \cite{gyevag2} improved this result for left-linear TRSs. They  showed the following.
Let R be a left-linear TRS over $sign(R)$,  and let $\S=\{\,g,\sharp\,\}\cup
sign(R),$ where $g\in\S_1-sign(R)$ and $\sharp\in\S_0-sign(R).$ Then $R$ is an  EP$\S$R-TRS  if and only if $R$ 
is an EPR-TRS.

In \cite{gill} Gilleron showed that 
for a TRS $R$ over $\S$ it is not decidable  whether
$R$ is a P$\S$R-TRS.
We may  naturally introduce 
the above concepts for string rewrite systems as well. 
Otto \cite{ott} has proved that 
a string rewrite system $R$ over the alphabet  $alph(R)$ of $R$ preserves $alph(R)$-recognizability if
 and only if $R$ preserves recognizability.
Otto \cite{ott} showed that it is not decidable for a   string rewrite system $R$ whether $R$ preserves  $alph(R)$-recognizability,
 and 
whether $R$   preserves recognizability. 
Hence it is not decidable for a linear TRS $R$ whether $R$ is a  PR-TRS, see \cite{ott}.

In spite of the undecidability results of Gilleron \cite{gill} and Otto \cite{ott}, 
 we know several  classes of EPR-TRSs. 
Gyenizse and V\'agv\"olgyi \cite{gyevag} generalized the concept of a 
semi-monadic TRS \cite{coq} introducing the concept of a 
generalized semi-monadic TRS (GSM-TRS for short). They showed that 
each 
linear GSM-TRS  $R$  is an EPR-TRS.  
Takai, Kaji, and Seki \cite{tks} introduced 
finite path overlapping TRS's (FPO-TRSs). They  \cite{tks} showed that each 
right-linear FPO-TRS  $R$  is an EPR-TRS. 
 They  \cite{tks} also showed that each
 GSM-TRS $R$ is an FPO-TRS.  Thus we get that 
 that each 
right-linear GSM-TRS  $R$  is an EPR-TRS. 
 V\'agv\"olgyi \cite{half} introduced the concept of a half-monadic TRS.
Each right-linear half-monadic TRS is an  FPO-TRS. 
Hence  each right-linear half-monadic TRS is an   EPR-TRS.
Using this result, V\'agv\"olgyi \cite{half} showed that 
 termination and convergence are decidable properties 
for right-linear half-monadic term 
rewrite systems. 
Takai, Seki, Fujinaka, and Kaji \cite{japanok} presented  an EPR-TRS which is not an
 FPO-TRS, see Example 1 in \cite{japanok}. 
 Takai, Seki, Fujinaka, and Kaji \cite{japanok}  introduced 
layered transducing term rewriting systems (LT-TRSs).
 They  \cite{japanok} showed that each 
right-linear LT-TRS  $R$  is an EPR-TRS.

We show that each terminating TRS is an EPRF-TRS.
We adopt the construction of  Salomaa \cite{sal}, Coquid\'e et al \cite{coq}, and 
Gyenizse and V\'agv\"olgyi \cite{gyevag}, when  showing
 that any left-linear GSM-TRS $R$ is an EPRF-TRS.  
We slightly modify  the proofs of the decision results of
Gyenizse and V\'agv\"olgyi \cite{gyevag} when we  show the following decidability results.

(1)
Let $R$ be an EPRF-TRS over $\S$, and let
 $p, q\in \ts(X)$. Then it is decidable whether  $p\tred R q$.

(2) Let $R$ be an EPRF-TRS over $\S$,
 and let
 $p, q\in \ts(X)$. Then it is decidable whether  there exists a 
tree $r\in \ts(X)$ such that $p\tred R r$ and $q \tred R r$.

(3) Let $R$ be  a confluent EPRF-TRS over $\S$,
 and let
$p, q\in \ts(X)$. Then it is decidable   whether   $p\tthue R q$.

(4)  For an EPRF-TRS  $R$, it is decidable whether
 $R$ is locally confluent.

 (5)
Let $R$ be  an EPRF-TRS, and let $S$ be a TRS over $\S$.
 Then it is decidable whether $\tred {S}\subseteq \tred {R}$.

(6)
Let $R$ and $S$  be  EPRF-TRSs. 
Then it is decidable which 
one of the following four mutually excluding conditions hold.

(i) $\tred {R}\subset  \tred {S}$,

(ii) $\tred {S}\subset  \tred {R}$,

(iii) $\tred {R}=  \tred {S}$,

(iv) $\tred {R}\Join \tred {S}$,

\noindent where ``$\Join$\,'' stands for the incomparability relationship.

(7)
Let $R$ be an  EPRF-TRS.  Then 
it is decidable whether 
$R$ is left-to-right minimal.
(A TRS $R$ is left-to-right minimal if for each 
rule $l\rightarrow r$ in $R$,
$\tred {R- \{\,l\rightarrow r\,\}}\subset \tred R$.)

(8)
Let $R$ and $S$  be TRSs such that $R\cup R^{-1}$ and 
$S\cup S^{-1}$ are  EPRF-TRSs. 
Then it is decidable which 
one of the following four mutually excluding conditions holds.

(i) $\tthue {R}\subset  \tthue {S}$,

(ii) $\tthue {S}\subset \tthue {R}$,

(iii) $\tthue {R}=  \tthue {S}$,

(iv) $\tthue {R}\Join \tthue {S}$.

F\"ul\"op's \cite{ful} undecidability results  
on deterministic top-down tree transducers
 simply imply the following.
Each of the following questions is 
undecidable for any convergent  left-linear  EPRF-TRSs
$R$ and $S$ over a
 ranked alphabet $\Omega$, for any recognizable tree language 
$L\subseteq T_\Omega$
given by a tree automaton over $\Omega$ recognizing $L$. Here 
 $\Gamma\subseteq \Omega$ is the smallest ranked alphabet for which
$R(L)\subseteq T_\Gamma$. Furthermore, the set of $R$-normal forms of the trees in $L$ is denoted by 
$R(L)$.

(i) Is $R(L)\cap S(L)$ empty?

(ii) Is $R(L)\cap S(L)$ infinite?

(iii) Is $R(L)\cap S(L)$ recognizable?

(iv) Is $T_\Gamma - R(L)$ empty?

(v) Is $T_\Gamma - R(L)$  infinite?

(vi) Is $T_\Gamma - R(L)$  recognizable?

(vii) Is $R(L)$ recognizable?

(viii) Is $R(L)=S(L)$?

(ix) Is  $R(L)\subseteq S(L)$?

F\"ul\"op and Gyenizse \cite{fulgye} showed that 
it is undecidable for a tree function induced by a 
deterministic homomorphism whether  it is injective.
Hence  for any convergent left-linear   EPRF-TRS $R$
over a ranked alphabet $\S$,  and  any recognizable tree language  $L\subseteq \ts$, it is undecidable whether  the tree function 
$\tred R\cap (L\times R(L))$ is injective.

Finally we show the following.
Let $R$ be a linear collapse-free EPRF-TRS
and   $S$ be a linear collapse-free EPR-TRS
 over the disjoint ranked alphabets 
$sign(R)$ and $sign(S)$, respectively. 
Then $R\oplus S$ is a   linear collapse-free  EPR-TRS.

This paper is divided into seven sections. In Section \ref{felleg}, we recall 
the necessary notions and notations. In Section \ref{zold}, we study TRS classes preserving recognizability.
In Section \ref{harmadik}, we show that left-linear 
GSM-TRSs are EPRF-TRSs.  In Section \ref{negyedik}, we illustrate  the  constructions presented  in Section \ref{harmadik}
by an example. 
 In Section \ref{marci}, we show various decidability and undecidability results on 
PRF-TRSs and EPRF-TRSs. 
Finally, in Section \ref{geza}, we present our concluding remarks,
and some open problems.

\section{Preliminaries}\label{felleg}

We recall and invent some notations, basic definitions and terminology which
will be used in the rest of the paper. Nevertheless the reader is assumed to be
familiar with the basic concepts of term rewrite systems and of tree language theory
(see, e.g. Baader, Nipkow\cite{baanip}, G\'ecseg, Steinby
\cite{gecste}, \cite{gecste2}). 

\subsection{Terms}
The cardinality of a set $A$ is denoted by $|A|$.
We denote by $\rho^{-1}$
the inverse of  a   binary relation $\rho$. 
The composition of  binary relations $\rho$ and $\tau$
is denoted by $\rho \c \tau$.

The set of nonnegative integers is denoted by $N$, and $N^*$ stands
for the free monoid generated by $N$ with empty word
 $\lambda$ as identity element. 
 For a word $\alpha\in N^*$, 
$length(\alpha)$ stands for the length of $\alpha$.

A ranked alphabet is a finite set $\S$ in which every symbol has a unique rank
in $N$. For $m\geq 0$, $\S_m$ denotes the set of all elements of $\S$ which
have rank $m$. 
The elements of $\S_0$ are called constants.
We assume that all ranked alphabets $\S$ and $\Delta$ that we consider
have the following property.
If $\s \in \S_i$,
 and $\s \in \Delta_j$, then $i=j$. In other words, $\s$ has the same rank in 
$\S$ as in $\Delta$.

For a set of variables $Y$ and a ranked alphabet $\S$, 
$T_\S(Y)$ denotes the set of $\S$-terms (or $\S$-trees) over $Y$. 
$\ts(\emptyset)$ is written as $\ts$. A term $t\in \ts$ is called a
ground term. 
A tree $t\in \ts(Y)$ is linear if any variable of 
$Y$ occurs at most once in $t$. 
We specify a countable set
$X=\{\,x_1,x_2,\ldots\,\}$ of variables which will be kept fixed in this paper.
Moreover, we put  $X_m=\{\,\seq x1m\,\}$, for $m\geq 0$. Hence $X_0=\emptyset$.

For any $m \geq 0$, 
we distinguish a subset $\bar{T}_\Sigma(X_m)$ of $\ts(X_m)$ as follows:
a tree $t\in T_\Sigma(X_m)$ is in $\bar{T}_\Sigma(X_m)$ if and only if
each variable in $X_m$ appears exactly once in $t$.

For a term $t\in \ts (X)$, the height
$height(t)\in N$, 
the set of variables $var(t)$ of $t$, 
and the set of positions $POS(t)\subseteq N^*$ are  defined in the usual way.
For each $t\in T_\S(X)$ and $\alpha\in POS(t)$, 
$t/\alpha\in \ts(X)$ is the subterm of $t$ at $\alpha$. 
For a tree $t\in\ts (X)$, $sub(t)$ denotes the
 {\em subtrees} of $t$. 
For a tree language $L\subseteq \ts$, the set $sub(L)$ of subtrees of elements of $L$ is defined by the
equality
$sub(L)=\bigcup(\, sub(t)\mid t\in L\,)$.

For $t\in\ts$, $\alpha\in POS(t)$, and $r\in\ts$, we define
$t[\alpha\leftarrow r]\in\ts$ as follows.

(i) If $\alpha=\lambda$, then $t[\alpha\leftarrow r]=r$.

(ii) If $\alpha=i\beta$, for some $i\in N $ and $\beta\in N^*$,
then $t=f(\seq t1m)$ with $f\in\S_m$ and $\les 1im$.
 Then $t[\alpha\leftarrow r]
=f(t_1,\ldots,t_{i-1},t_i[\beta\leftarrow r],t_{i+1},\ldots,t_m)$.

Let $\S$ be a ranked alphabet. Let $f\in \S_1$, $t\in \ts$ be arbitrary.
The tree $f^k(t)\in \ts$, $k\geq 0$, is defined by recursion:
$f^0(t)=t$, and $f^{k+1}(t)=f(f^k(t))$ for $k\geq 0$.

A substitution 
is a mapping $\s: X\rightarrow\ts(X)$ which is different from the
identity only for a finite subset $Dom(\s)$ of $X$.
For any substitution $\s$ with $Dom(\s)\subseteq X_m$, $m\geq 0$, 
the term $\s(t)$ is produced from
$t$ by replacing each occurrence of  $x_i$ with $\s(x_i)$ for $1\leq i \leq m$.
For any trees $t\in\bar{T}(X_k)$, $ t_1,\ldots
,t_k\in\ts(X)$ and for the substitution $\s$ with $Dom(\s)\subseteq X_k$
 and
$\s(x_i)=t_i$ for $i=1,\ldots,k$, we denote the term $\s(t)$ 
by $t[t_1,\ldots ,t_k]$ as well. 
Moreover, for any $k,m$ with $1\leq m\leq k$, for any 
tree $t\in T_\Sigma(\{\, x_m , \ldots , x_k\,\})$ and 
for any  substitution $\s$ with
$\s(x_m)=t_m, \ldots, \s(x_k)=t_k$,
we denote $\s(t)$ also  by $t[x_m\leftarrow t_m, \ldots, 
x_k \leftarrow t_k]$.

Let $\S$ be a ranked alphabet and   let $u,  v\in \ts(X)$. 
The tree $u$ is a  supertree of $v$ if $u$ is linear and there is a substitution $\s$ such that 
$v=\s(u)$.

 We say that the pair $(l_1,r_1)$ is a variant 
of the pair  $(l_2, r_2)$ if there is a substitution 
$\s : X\rightarrow X$ such that 

(i) $\s(l_2)=l_1$ and $\s(r_2)=r_1$, and that 

(ii) for all $x_i, x_j \in var(l_2) \cup var(r_2)$, 
$\s(x_i)=\s(x_j)$ implies that $x_i=x_j$.

\noindent
For the concept of a unifier and a most general unifier (mgu), see \cite{baanip}.

\subsection{TRSs}

Let $\S$ be a ranked alphabet. Then a term rewrite system (TRS) $R$ over $\S$ is a finite
subset of $\ts(X)\times\ts(X)$ such that for each $(l,r)\in R$, each 
variable of
$r$ also occurs in $l$. Elements $(l,r)$ of $R$ are 
called rules and are denoted
by $l\rightarrow r$. 
Furthermore, $sign(R)\subseteq \S$ is the
ranked alphabet consisting of all symbols appearing in the rules of $R$.

Let $R$ be a TRS over $\S$.
Given any two terms $s$ and $t$ in $\ts(X)$ and an position 
$\alpha\in POS(s)$, 
we say that $s$ rewrites to $t$ at $\alpha$
and denote this by $s\red R t$ if there is some pair $(l, r)\in R$ and a
substitution $\s$ such that
 $s/\alpha=\s(l)$ and $t=s[\alpha\leftarrow \s(r)]$. 
Here we also say that $R$ rewrites $s$ to $t$ applying the rule 
$l\rightarrow r$ at $\alpha$.

 Reachability, joinability, termination, confluence, 
local confluence, convergence
are defined in the usual way, see \cite{baanip}.

We say that  a TRS $R$ is collapse-free if there is no rule
$l\rightarrow r $ in $R$ such that $l\in X$ or $r\in X$.

A left-linear (linear, resp.) TRS is one in
which no variable occurs more than once on any left-hand side 
(right-hand side and left-hand side, resp.).
A ground TRS is one of which all rules are ground
 (i.e. elements of $\ts\times\ts$).

A TRS 
is monadic if each
left-hand side is of height at least $1$ and each right-hand side is of 
height at most $1$.
A TRS is called right-ground
 if each right-hand side is ground.
A TRS $R$ over $\S$
is {\em murg} if $R$ is the union of a monadic  TRS and a right-ground  TRS over $\S$.
Obviously, each monadic TRS is murg, and each  right-ground  TRS is murg. 
For the concept of a finite path overlapping TRS  (FPO-TRS) see \cite{tks}. 
For the concept of a  layered transducing TRS (LT-TRS), see \cite{japanok}.

Let $R$ be a rewrite system over $\S$.

(a) $R$ is left-to-right minimal if for each rule $l\rightarrow r$ in $R$,
$\tred {R- \{\,l\rightarrow r\,\}}\subset \tred R$.

(b) $R$ is  left-to-right 
ground  minimal if for each rule $l\rightarrow r$ in $R$,
$\tred {R- \{\,l\rightarrow r\,\}}\cap( \ts \times \ts)\subset \tred R
\cap( \ts \times \ts)$.

The set of all ground terms that are irreducible 
for a TRS $R$ is denoted by $IRR(R)$.

Let $R$ be a convergent TRS over $\S$, and let $p\in \ts(X)$.
It is well known that there exists exactly one term $t\in \ts(X)$ 
irreducible for $R$ such that $p\tred R t$.
 We call $t$ the $R$-normal form of $p$. We denote $t$ by $p\hspace{-1.5mm}\downarrow_R$. 
Let $L\subseteq \ts$. 
The set of $R$-normal forms of the trees in
the tree language $L$ is denoted by 
$R(L)$. It should be clear that  $R(L)=R^*(L)\cap IRR(R)$.

For the concept of a critical pair, see \cite{baanip}. 

Let $R$ and $S$ be rewrite systems over the disjoint ranked alphabets 
$\S$ and $\D$, respectively. Then the disjoint union 
$R\oplus S$ of $R$ and $S$  is the rewrite system 
$R\cup S$ over the ranked alphabet $\S\cup \D$.
Let ${\bf C}$ be  a class of rewrite systems, let ${\bf C}$ be 
closed under disjoint union.
 A property $P$ is modular for 
${\bf C}$ if for any $R,S \in {\bf C}$ over disjoint ranked alphabets, 
$R\oplus S$ has the property ${\cal P}$
if and only if 
both $R$ and $S$ have the property ${\cal P}$.

\subsection{Tree Languages}

Let $\S$ be a ranked alphabet, a bottom-up tree automaton 
(bta) over 
$\S$ is a quadruple ${\cal A}=(\S,A,R, A_f)$,
 where  $A$ is a finite set of states of rank $0$, 
$\S \cap A=\emptyset$, 
$A_f(\subseteq A)$
is the set of final states, 
$R$ is a finite set of rules of the following
 two types:

(i) $ \delta(e_1,\ldots, e_n)\rightarrow a$ with $n\geq 0$, $\delta\in \S_n$,
$\seq a1n, a\in A$.

(ii) $ a \rightarrow a'$ with $a,a'\in A$ ($\lambda$-rules).

\noindent
We consider $R$ as a ground TRS over $\S\cup A$. 
The tree 
language recognized by $\cal A$ is $L({\cal A})=\{\,
t\in \ts\mid (\exists a\in A_f) \;t\tred {R} a\}$.
A tree language $L$ is recognizable if there
exists a bta $\cal A$ such that $L({\cal A})=L$
(see \cite{gecste}). 

The bta 
${\cal A}=(\S,A,R, A_f)$ 
is deterministic if $R$ has no $\lambda$-rules and $R$ has no two rules
with the same left-hand side. 
\begin{df}{\em 
Let $\S$ be a ranked alphabet, and let $L\subseteq \ts$ be a finite tree
language. We define the {\em  the fundamental bta}  
${\cal A}=(\S,A,R, A_f)$  of $L$ as follows.
 
$A=\{\, \langle p \rangle \mid \mbox{ there is }q \in L\mbox{ such that }p\in sub(q)\, \}$.

$R=\{\,  \delta(\langle p_1 \rangle, \ldots, \langle p_m \rangle)\rightarrow \langle p\rangle  \mid \delta\in \S_m, m \geq 0, 
\langle p_1 \rangle, \ldots, \langle p_m \rangle, \langle p\rangle \in A$,  and 

\hspace{5.5 cm} $p=\delta(p_1, \ldots, p_m)\, \}$.
 
$A_f=\{\, \langle p \rangle\mid p\in L\, \}$.

}
\end{df}

\begin{lem}\label{tokaj}
Let $\S$ be a ranked alphabet, and let $L\subseteq \ts$ be a finite tree
language. Let ${\cal A}=(\S,A,R, A_f)$ be the fundamental bta of $L$. Then
   ${\cal A}$ is deterministic, and $L({\cal A})=L$.
\end{lem}
\pr By the definition of $R$,  ${\cal A}$ is deterministic. 
It is not hard to see that
for any  $t\in sub (L)$ and $a \in A$, 
$t \tred A a$ if and only if $a=\langle t\rangle $.
By the definition of $A_f$, $L({\cal A})=L$. 

\hfill $\b$


\section{TRSs Preserving Recognizabiliy}\label{zold}

Let $\S$ be a ranked alphabet, let
$R$ be a TRS over $\S$, and let $L$ be a tree language over $\S$. 
Then $R^*_\S(L)=\{\,p\mid
q\tred R  p \mbox{ for some }q\in L\,\}$ is the set of descendants of trees in
$L$. When $\S$ is apparent from the context, we simply write $R^*(L)$
 rather than $R^*_\S(L)$.

For the concept of a  P$\S$R-TRS,  a P$\S$RF-TRS, a PR-TRS,  a PRF-TRS, 
an EP$\S$R-TRS,  an EP$\S$RF-TRS, an EPR-TRS,  and an EPRF-TRS, see the Introduction.

\begin{tet} Each terminating TRS is an EPRF-TRS.
\end{tet} \pr Let $R$ be a terminating TRS over $\S$ and let $L\subseteq \ts$ be a finite language.
Let $t \in \ts$ be arbitrary. We now show that  $R^*(\{\, t\,\})$ is finite. On the contrary, assume that
 $R^*(\{\, t\,\})$ is infinite. Then 
$t$  starts an infinite reduction sequence
$t=t_0\red R t_1 \red R t_2 \red R t_3 \red R \cdots$
 by K\"onig's lemma. Hence $R$ is not terminating, which is a contradiction.

 Thus $R^*(L)$ is finite and hence recognizable.
We  compute $R^*(L)$ as follows. Let $W=L$. While there is $q\in \ts-W$ such that $p \red R q$ for some $p\in W$
we add $q$ to $W$.  Since $R^*(L)$ is finite, we stop.
When we stop we have  $W=R^*(L)$. Then we can construct a bta $\cal C$ over $\S$ such that $L({\cal C})=W$. 
\hfill $\b$


\begin{sta}\label{pamir}
There is a left-linear monadic TRS $R$ over a ranked alphabet 
$\S$ such that  
$R$ is an EP$\S$RF-TRS,  and that
$R$ is not a P$\S$R-TRS. 
\end{sta} 
\pr Let $\S=\S_0\cup\S_2$, $\S_0=\{\, \sharp\,\}$, 
and $\S_2=\{\, f\,\}$. Let $R$ over $\S$ consist of the rule
$f(x_1, \sharp) \rightarrow f(x_1, x_1)$. Observe that $\S=sign(R)$. We obtain by direct inspection  that
 $R$  is a left-linear monadic TRS $R$ over 
$\S$. 
Let $L\subseteq \ts$ be an arbitrary finite tree language. For any trees
$p, q\in \ts$, if $p \red R q$, then $height(p)= height(q)$. Hence $R^*(L)$ is finite. Thus $R^*(L)$ is recognizable.
We  compute $R^*(L)$ as follows. Let $W=L$. While there is $q\in \ts-W$ such that $p \red R q$ for some $p\in W$
we add $q$ to $W$. When we stop we have  $W=R^*(L)$. We can construct a bta $\cal C$ over $\S$ such that $L({\cal C})=W$. 

Then  we define the $n$th left comb
$left_n$ for $n\geq 0$, as follows.

(i)
$left_0=\sharp$, and 

(ii)
for each $n\geq 0$, 
$left_{n+1}=f(left_n, x_{n+1})$.

\noindent 
Let $$L=\{\,  left_n\mid n\geq 0\, \}\, .  $$
Then $L$ is a recognizable tree language.
 Furthermore, 
for any $p\in R^*(L)$ and  $s\in sub(p)$, if $s=f(t_1, t_2)$, then $t_2=\sharp$ or 
$height(t_1)=height(t_2)$. 
For each $n\geq 0$, $f(left_n, left_n)\in  R^*(L)$. Assume that $R^*(L)$ is a recognizable tree language. 
Similarly to  the proof of  the pumping lemma for recognizable tree languages, one can show the following.
There are 
$0\leq i< j$ such that  $f(left_i, left_j)\in  R^*(L)$.
However, $height (left_i)=i < j =height( left_j)$. This contradicts our observation on $ R^*(L)$. 

\hfill $\b$

\begin{sta}\label{alfold}
There is an LT-TRS  $R$ over a ranked alphabet 
$\S$ such that  $R$ is not a P$\S$RF-TRS. 
\end{sta} 
\pr Let $\S={\cal F} \cup {\cal Q}=\S_0\cup \S_1$, $\S_0=\{\, \sharp\,\}$, 
$\S_1=\{\, f, q \,\}$,   ${\cal F}=\{\, \sharp, f\,\}$, and ${\cal Q}=\{\, q\, \}$. 
Let TRS $R$ consist of the rules

$f(q(x_1)) \rightarrow  q(f(x_1))$, 

$\sharp\rightarrow  q(f(\sharp))$. 

\noindent
Then $R$  is  an LT-TRS. 
It is not hard to see that 
$$R^*(\{\, \sharp\, \})\cap \{\, f^k(q^m(\sharp))\mid k, m\geq 0\, \}\,
= \{\, f^k(q^k(\sharp))\mid k\geq 0\, \}\, .  $$
Here  $ \{\, f^k(q^m(\sharp))\mid k, m\geq 0\, \}\,   $ is a recognizable tree language, and 
 $ \{\, f^k(q^k(\sharp))\mid k\geq 0\, \}\,   $ is not a recognizable tree language.
The intersection of two recognizable tree languages is also a recognizable tree language.
Hence $R^*(\{\, \sharp\, \})$ is not a recognizable tree language.

\hfill $\b$


\begin{tet}
There is a  ranked alphabet $\S$ and a murg TRS $R$ over $\S$ such that $R$ 
is not a P$\S$RF-TRS. 
\end{tet}\pr
Let $\S=\S_0\cup \S_1\cup\S_2$, $\S_0=\{\, \sharp, \$, \flat\,\}$, 
$\S_1=\{\, f\,\}$, and $\S_2=\{\,g, h \,\}$. Let the  TRS $R$
over $\S$ consist of the  rules

$\sharp \rightarrow f(\sharp)$,

$\sharp \rightarrow \flat$,

$\$ \rightarrow f(\$)$,

$\$ \rightarrow \flat$,

$g(x_1, x_1)\rightarrow h(x_1, x_1)$,

\noindent 
Consider a tree $ h(t_1, t_2)\in R^*(g(\sharp, \$))$ where $t_1, t_2 \in  \ts$.
Then 

$g(\sharp, \$) \tred R g(f^k (\sharp), f^k (\$))
\tred R g(f^k (\flat), f^k (\flat)) \red R h(f^k (\flat), f^k (\flat))= 
h(t_1, t_2)$

\noindent 
holds for some $k\geq 0$.
Hence 

\noindent 
$R^*(\{\, g(\sharp, \$)\, \})\cap \{\, h(t_1, t_2)\mid \, t_1, t_2\in \ts\}=
 \{\, h(f^k(\flat), f^k(\flat))\mid k\geq 0\}$. 

\noindent 
It is well known  that the intersection of any two recognizable tree languages is a recognizable tree language.
Observe that $ \{\, h(t_1, t_2)\mid \, t_1, t_2\in \ts\}$ 
is a recognizable tree language, and   $\{\, h(f^k(\flat), f^k(\flat))\mid k\geq 0\}$
is not a recognizable tree language. 
Thus $R^*(\{\, g(\sharp, \$)\, \})$ is not a recognizable tree language.

\hfill $\b$

With an arbitrary Post Correspondence System (PCS) ${\langle{\bf w}, {\bf z}\rangle}$,  
V\'agv\"olgyi \cite{vag} associated a ranked alphabet $\S$,  containing the distinguished   
nullary symbol $\#\in \S_0$,  
 and a TRS  $R$  over 
$\S$. 
V\'agv\"olgyi \cite{vag} showed the following results.

\begin{sta}\label{tizenketto} If PCS ${\langle{\bf w}, {\bf z}\rangle}$ has a solution, then 
$R$ is an  EP$\S$R-TRS. 

\end{sta}
\begin{sta}\label{torta} If PCS ${\langle{\bf w}, {\bf z}\rangle}$ has no solution, then
$R^*(\{\, \#\, \})$ is not a recognizable tree language over $\S$.
\end{sta}
Statement \ref{torta} implies the following statement.

\begin{sta}\label{tizennegy} If PCS ${\langle{\bf w}, {\bf z}\rangle}$ has no solution, then
 $R$ is not a P$\S$RF-TRS. 
\end{sta}

\noindent 
The following result is a simple consequence of Statements \ref{tizenketto} and \ref{tizennegy}.

\begin{sta}\label{tizenhat} PCS ${\langle{\bf w}, {\bf z}\rangle}$ has a solution if and only if 
$R$ is an EP$\S$RF-TRS if and only if TRS $R$ is a P$\S$RF-TRS.

\end{sta}

\noindent 
Statement \ref{tizenhat} implies the following result.
\begin{prop}\label{alpha} The following problem is undecidable:

{\em  Instance:} A  
 murg TRS $R$   over  a ranked alphabet $\S$.

{\em Question:} Is $R$ a P$\S$RF-TRS?
\end{prop}


We now recall  the notion of a GSM-TRS, see \cite{gyevag}.

\begin{df}\label{GSM}{\em
Let $R$ be a TRS over $\S$. We say that 
$R$ is a generalized semi-monadic TRS
(GSM-TRS for short) if there is no rule
$l\rightarrow r$ in $R$ with $l\in X$  and the following holds.
For any rules $l_1\rightarrow r_1$ and $l_2\rightarrow r_2$ in $R$,
 for any positions $\alpha\in POS(r_1)$ and $\beta\in POS(l_2)$, 
and for any supertree $l_3\in \ts(X)$ of $l_2/\beta$
with $var(l_3)\cap var(l_1)=\emptyset$,
 if

(i) $\alpha=\lambda \mbox{ or }\beta=\lambda,$

(ii) $r_1/\alpha$ and $l_3$ are  unifiable,
and  

(iii) $\s$ is a most general unifier of $r_1/\alpha$ and $l_3$, 

\noindent then

 (a) $l_2 /\beta\in X$ or 

(b)  for each  $\gamma\in POS(l_3)$,
if $l_2/\beta\gamma\in X$, then $\s(l_3/\gamma)\in X \cup \ts$.
}
\end{df}
Notice that Condition (a) implies that $l_3\in X $.

\begin{exa}{\em 
Let $\S=\S_0\cup \S_1\cup\S_3$, $\S_0=\{\, \sharp\,\}$, 
$\S_1=\{\, f\,\}$, and $\S_3=\{\, g\,\}$. Let the  TRS $R$
over $\S$ consist of the  rule
$$g(x_1, x_2, \sharp)\rightarrow f(g(x_1, \sharp, x_1))\, .$$
We obtain by direct inspection that $R$ is left-linear  GSM-TRS.}
\end{exa}

\noindent
The proof of the following result is  straightforward.
\begin{obs} Each murg TRS is a GSM-TRS as well.
\end{obs}

\noindent
Gyenizse and V\'agv\"olgyi \cite{gyevag} observed that 
 F\"ul\"op's \cite{ful} undecidability results  
on deterministic top-down tree transducers
 simply imply the following. 
\begin{sta}\label{zoli} {\em \cite{gyevag}}
Each of the following questions is 
undecidable for any convergent  left-linear  
GSM-TRSs 
$R$ and $S$ over a
 ranked alphabet $\Omega$, for any recognizable tree language 
$L\subseteq T_\Omega$
given by a tree automaton over $\Omega$ recognizing $L$,
 where $\Gamma\subseteq \Omega$ is the smallest ranked alphabet for which
$R(L)\subseteq T_\Gamma$.

(i) Is $R(L)\cap S(L)$ empty?

(ii) Is $R(L)\cap S(L)$ infinite?

(iii) Is $R(L)\cap S(L)$ recognizable?

(iv) Is $T_\Gamma - R(L)$ empty?

(v) Is $T_\Gamma - R(L)$  infinite?

(vi) Is $T_\Gamma - R(L)$  recognizable?

(vii) Is $R(L)$ recognizable?

(viii) Is $R(L)=S(L)$?

(ix) Is  $R(L)\subseteq S(L)$?

\end{sta}

\begin{tet}
There is an  FPO-TRS $R$ such that $R$ is not an PRF-TRS.

\end{tet}
\pr
Let $\S=\S_0\cup \S_1 \cup\S_2$, $\S_0=\{\, \$\,\}$, 
$\S_1=\{\, d, g \,\}$, $\S_2=\{\, f\,\}$. Let the  TRS $R$
over $\S$  consist of the following rules.

$\$ \rightarrow d(\$)$

$g(d(x_1))\rightarrow f(g(x_1), d(x_1))$,

$g(\$) \rightarrow \$$,

$f(\$, x_1) \rightarrow \$$,

$f(\$, x_1) \rightarrow h(x_1, x_1)$.

\noindent
By direct inspection of $R$, we get that  $R$ is an FPO-TRS. We now study the set
$R^*(\{\, \{\, g(\$)\, \}\, \})\cap \{\, h(t_1, t_2)\mid t_1, t_2 \in \ts\, \}$. 
Assume that 

$g(\$)=u_0\red R u_2 \red R u_3\red R \cdots \red R u_{k-1}\red R u_k=h(t_1, t_2)$ 

\noindent for some 
$k\geq 1$ and $t_1, t_2 \in \ts$. 
We iterate application of 
the first and the second rules.
 We can change the order of  applications of the first and 
 second rules. Then we apply the third rule. Then we we apply the fourth rule finitely many times.
We apply the fifth rule in the $k$th step,  and hence $u_k =f(\$, t_1)$, and $t_1=t_2$.
Thus   we obtain
the following reduction sequence for some $n\geq 1$:

$g(\$)\red R g(d(\$)) \red R g(d^2(\$)) \red R \cdots \red R g(d^n(\$))\red R $

$f(g(d^{n-1}(\$), d^n(\$) )\red R f(  f(g(d^{n-2}(\$),d^{n-1}(\$) ),  d^n(\$) )
\red R $

$ f(  f(   f(g(d^{n-3}(\$) ,d^{n-2}(\$) ), d^{n-1}(\$) )),  d^n(\$))\red R    \cdots \red R$

$f( \ldots f(f(f(g(\$), d(\$)), d^2(\$)), d^3(\$)), \ldots,  d^n(\$))\red R $

$f( \ldots f(f(f(g(\$), d(\$)), d^2(\$)), d^3(\$)), \ldots,  d^n(\$))\red R $

$f( \ldots f(f(f(\$, d(\$)), d^2(\$)), d^3(\$)), \ldots,  d^n(\$))\red R $

$f( \ldots f(f(\$, d^2(\$)), d^3(\$)), \ldots,  d^n(\$))\red R $

$f( \ldots f(\$, d^3(\$)), \ldots,  d^n(\$))\red R \cdots \red R
f(\$, d^n(\$))\red R  h(d^n(\$), d^n(\$))$.

\noindent
In the light of the  above reduction sequence, one can show that 
 
$R^*(\{\, \{\, g(\$)\, \}\, \})\cap \{\, h(t_1, t_2)\mid t_1, t_2 \in \ts\, \}=
 \{\, h(d^n(\$), d^n(\$))\mid n\geq 1\, \}$.

\noindent
Since  $ \{\, h(t_1, t_2)\mid \, t_1, t_2\in \ts\}$ 
is a recognizable tree language, and   $  \{\, h(d^n(\$), d^n(\$))\mid n\geq 1\, \}$
is not a recognizable tree language, we get that
$R^*(\{\, g(\$)\, \})$ is not a recognizable tree language.

\hfill $\b$

\section{Main Results}\label{harmadik}
We now show that each left-linear GSM-TRS is an   EPRF-TRS.  

\begin{tet}\label{fotetel}
Each left-linear GSM-TRS is an  EPRF-TRS. 
\end{tet}
\pr
Let  $R$ be a left-linear GSM-TRS
over some ranked alphabet $\S$.   Moreover, 
let $L$ be a finite tree language over $\S$.
Via a series of  Lemmas we  show that 
$R^*(L)$ is recognizable. To this end, 
 we construct a bta 
$\cal C$ over $\S$ such that $L({\cal C})=R^*(L)$.
Our construction is illustrated by
an example in Section 4.

Let $E$ be the set of all ground terms $u$ over $\S$ such that 
there are rules $l_1\rightarrow r_1$ and $l_2\rightarrow r_2$ in $R$,
and there are  positions $\alpha\in POS(r_1)$ and $\beta\in POS(l_2)$, 
and there is a  supertree $l_3\in \ts(X)-X$ of $l_2/\beta$
with $var(l_3)\cap var(l_1)=\emptyset$ 
 such that 

(i) $\alpha=\lambda \mbox{ or }\beta=\lambda,$

(ii) $r_1/\alpha$ and $l_3$ are  unifiable,
and  

(iii) $\s$ is a most general unifier of $r_1/\alpha$ and $l_3$, 
  and 

(iv) there is a position   $\gamma\in POS(l_3)$
such that  $l_2/\beta\gamma\in X$ and  $\s(l_3/\gamma)\in \ts$
and $u=\s(l_3/\gamma)$.

\noindent 
It should be clear that $E$ is finite and is effectively constructable.

 Moreover, without loss of generality we may
assume that for each rule $l\rightarrow r$ in $R$, $l\in \bar{T}_\S(X_n)$
for some $n\geq 0$.
Let

\vspace{2mm}
$D=sub(L)\cup $

$\{\, p[e_1, \ldots, e_n] \mid n\geq 0, p\in \ts(X_n), 
e_1, \ldots, e_n\in sub(L\cup E), 
p \mbox{ is a subtree} $

\hspace {25 mm}$\mbox{ of the right-hand side }r \mbox{ of some rule }
 l\rightarrow r 
\mbox{ in } R\,\}\, .$

\vspace{2mm}

\noindent
Apparently, $sub(E) \subseteq D$. Hence $sub(L\cup E) \subseteq D$. 

Let
${\cal A}=(\S, A, S_A, A')$ be  the fundamental bta of $L$. 
Recall that ${\cal A}$ is  a deterministic
 bta over $\S$ such that $L({\cal A})=L$.
 Let 
${\cal B}=(\S, B, S_B, B')$ be  the fundamental bta of $D$. 
Recall that ${\cal B}$ is  a deterministic
 bta over $\S$ such that $L({\cal B})=D$.
By the definition of $D$, we have $A \subseteq B$ and 
\begin{equation}\label{alma}
S_A \subseteq S_B\, . 
\end{equation}
For each $i\geq 0$, consider the bta
${\cal C}_i=(\S, B, S_i, A')$, where
$S_i$ is defined by recursion on $i$
(for an example see Section 4).
Let 
\begin{equation}\label{korte}
S_0= S_B\, . 
\end{equation}
Then 
${\cal C}_0=(\S, B, S_B, A')$.  
Let us assume that $i\geq 1$ and we have defined the set $S_{i-1}$.
Then we define $S_{i}$ as follows. 

(a) $S_{i-1}\subseteq S_{i}$.

(b) For any rule $l\rightarrow r$ in $R$ with $n\geq 0$, 
$l\in \bar{T}_\S(X_n)$, for all $e_1, \ldots, e_n\in sub(L\cup E)$,
 if $l[\langle e_1 \rangle, \ldots,\langle e_n\rangle]\tred {S_{i-1}} c$
 for some 
$c\in B$,  then we put the rule

\noindent $\langle r[e_1, \ldots, e_n]
\rangle \rightarrow c$ 
in $S_{i}$.

\noindent
By (\ref{alma}) and (\ref{korte}),
 we have   
\begin{equation}\label{martzi}
S_A \subseteq S_{0}\, .
\end{equation}
It should be clear that there is an integer 
$M\geq 0$ such that $S_M=S_{M+1}$. Let $M$ be the least integer such that
$S_M=S_{M+1}$. Let ${\cal C}={\cal C}_M$. 
Let $S=S_M$, and 
from now on we write
${\cal C}=(\S, B, S, A')$, rather than ${\cal C}_M=(\S, B, S_M, A')$.

Our aim is to show that $R^*(L)=L({\cal C})$. To this end,
first we show five preparatory lemmas, then the inclusion $L({\cal C})
\subseteq R^*(L)$, then again five preparatory lemmas, and finally 
the inclusion $R^*(L)\subseteq L({\cal C})$.
\begin{lem}\label{pest}
$L=L({\cal C}_0)$.
\end{lem}
\pr By Lemma \ref{tokaj}, $L({\cal A})=L$. 
By the definition of $\cal B$ and ${\cal C}_0$,  we have 
$L({\cal C}_0)=L({\cal A})=L$. 

\hfill $\b$

\begin{lem}\label{sarosp}
For any $p\in \ts$ and  $r\in {T}_\S(X_n)$ with $n\geq 0$ and  $var(r)=X_n$, 
if $p\tred {S_0} \langle r[e_1, \ldots, e_n]\rangle$, 
 then $p=r[e_1, \ldots, e_n]$.
\end{lem}
\pr By direct inspection of the rules of $S_0$. 

\hfill $\b$

\noindent
The following statement is a simple consequence of Lemma
\ref{sarosp}.

\begin{lem}\label{kopaszka}
For any $p\in \ts$, $r\in T_\S(X_n)$, and $e_1, \ldots, e_n\in  sub(L\cup E)$,
if $p\tred {S_0} \langle r[e_1, \ldots, e_n]
\rangle$,  then $p=r[e_1, \ldots, e_n]$.
\end{lem}

\begin{lem} \label{csongrad}
For any $i \geq 1$, $p\in \ts$, $q,t\in T_{\S \cup B}$, $k\geq 1$,
and $v_1, \ldots , v_k\in T_{\S\cup B}$, if
\begin{equation} \label{fofasor}
p=v_1\red {S_0} v_2 \red {S_0} \ldots \red {S_0} v_k=q \red {S_i} t\, ,
\end{equation}
and ${\cal C}_i$ applies an 
$(S_i-S_{i-1})$-rule in the last step $q \red {S_i} t$
of {\em (\ref{fofasor})}, then there exists an $s\in \ts$ such that 
\begin{equation}\label{szarvas} 
s\red R p \mbox{ and } s \tred {S_{i-1}} t\, .
\end{equation}
\end{lem}
\pr
Let $\alpha$ be the position where 
${\cal C}_i$ applies an $(S_i-S_{i-1})$-rule

$$\langle r[e_1, \ldots, e_n]\rangle \rightarrow c
\mbox{ with }r\in T_\S (X_n), n\geq 0, \mbox{ and }
e_1, \ldots, e_n \in  sub(L\cup E)$$
in the last step $q \red {S_i} t$
of (\ref{fofasor}). Then 
$$q=u[\langle r[e_1, \ldots, e_n]\rangle ]\, ,$$ 
where $u\in \bar{T}_\S(X_1)$, $u/\alpha=x_1$. 
By Lemma \ref{kopaszka}, 
$$p=u[r[e_1, \ldots, e_n]]\, . $$
 Finally,
$t=u[c]$.
By (b) of 
the definition of $S_i$, $i\geq 1$, there is a rule $l \rightarrow r$
in $R$ with $l\in \bar{T}_\S(X_n)$, $n\geq 0$
such that 
$$l[\langle e_1\rangle, \ldots, \langle e_n\rangle]\tred {S_{i-1}} c\, . $$
Let $$s=u[l[e_1, \ldots , e_n]]\, .$$ Then 
$$s \red R p$$
and $$s =u[l[e_1, \ldots, e_n]]\tred {S_0} u[l[\langle e_1 \rangle, \ldots , 
 \langle e_n \rangle]]\tred {S_{i-1}} u[c]=t\, .$$
Hence (\ref{szarvas}) holds.
\hfill $\b$
\begin{lem}\label{mako}
For any $i\geq 0$, $ p\in \ts$, and $q\in T_{\S\cup B}$, 
if $p\tred {S_i} q$, then there is an $s\in \ts$ such that 
$$
s\tred R p \mbox{ and } s\tred {S_0} q\, .
$$
\end{lem}
\pr We proceed by induction on $i$. For $i=0$ the statement is trivial. 
Let us suppose that $i\geq 1$ and that we have shown the statement for $1,2,
\ldots, i-1$. Let
\begin{equation}\label{dupla}
p\tred  {S_i} q\, ,
\end{equation}
and let $m$ be the number of $(S_i-S_{i-1})$-rules applied by $C_i$ along 
(\ref{dupla}).
We show by induction on $m$ that 
\begin{equation}\label{kettos}
\mbox{ there is  $s\in \ts$ such that }
s\tred R p \mbox{ and } s\tred {S_0} q\, .
\end{equation}
If $m=0$, then $p\tred {S_{i-1}} q$ and hence by the
induction hypothesis on $i$,  (\ref{kettos}) holds.

Let us suppose that $m\geq 1$ and that for $0, 1, \ldots, m-1$, we have shown 
(\ref{kettos}). Let $p\tred {S_i} q$ where $\cal C$ applies $m$ 
$(S_i-S_{i-1})$-rules. 
Then there are integers $n,k$, $1\leq k\leq n$, and there are trees 
$t_1, t_2, u_1, u_2, \ldots, u_n\in T_{\S\cup B}$
 such that (I), (II), (III), and (IV) hold.

(I) $p=u_1\red {S_i} \ldots \red {S_i} u_k=t_1\red {S_i}u_{k+1}=t_2
\red {S_i} \ldots \red {S_i} u_n=q$.

(II)  along the reduction subsequence  
$p=u_1\red {S_i} \ldots \red {S_i} u_k=t_1$
of (I), ${\cal C}_i$ applies no $(S_i-S_{i-1})$-rule.

(III)
in the rewrite step $u_k\red{S_i} u_{k+1}$ ${\cal C}_i$
applies an $(S_i-S_{i-1})$-rule.

(IV)
along the reduction subsequence 
$t_2=u_{k+1}\red {S_i} \ldots \red {S_i} u_n=q  $
of (I), ${\cal C}_i$ applies  $m-1$ $(S_i-S_{i-1})$-rules. 

\noindent
By the induction hypothesis on $i$, there is a tree $s_1\in \ts$ such that 
\begin{equation}\label{tolna}
s_1 \tred R p \mbox{ and } s_1\tred {S_0} t_1\, .
\end{equation}
Hence
 $$s_1\tred {S_0} t_1 \red {S_i} t_2\, .$$
By Lemma \ref{csongrad}, there is a tree $s_2\in \ts$ 
such that \begin{equation}\label{baranya}
s_2\red R s_1 \mbox{ and } s_2 \tred {S_{i-1}} t_2\, .
\end{equation}
Hence there is $j\geq 0$ 
and there are $w_1, \ldots, w_j\in T_{\S\cup B}$ such that 
\begin{equation}\label{fekete}
s_2=w_1 \red {S_{i-1}} w_2 \red {S_{i-1}} \ldots \red {S_{i-1}} 
w_j=t_2=u_{k+1}\red {S_i} \ldots \red {S_i}u_n=q\, ,
\end{equation}
 and along (\ref{fekete}), ${\cal C}_i$ applies $m-1$ 
$(S_i-S_{i-1})$-rules. 
By the induction hypothesis on $m$, there is a tree 
$s_3\in \ts$ such that 
$$s_3 \tred R s_2 \mbox{ and } s_3 \tred {S_0} q\, .$$
Hence by (\ref{tolna}) and (\ref{baranya}),
$$s_3 \tred R s_2 \red R s_1 \tred R p\, .$$
Thus (\ref{kettos}) holds.
\hfill $\b$

\begin {lem}\label{elso}
$L({\cal C})\subseteq R^*(L)$.
\end{lem}
\pr Let $p\in L({\cal C})$.
Then $p\tred {S} b$ for some $b\in A'$. Hence by Lemma 
\ref{mako}, there is an $s\in \ts$ such that 
\begin{equation}\label{bakony}
s\tred R p \mbox{ and }s
\tred {S_0} b\, .
\end{equation}
 Hence $s\in L({\cal C}_0)$. By Lemma \ref{pest}, $s\in L$. Thus by
 (\ref{bakony}),    $p\in R^*(L)$.
\hfill $\b$

\vspace{3.mm}
Now we show the inclusion $R^*(L) \subseteq L({\cal C})$. To this end,
first we prove five lemmas.

\begin{lem}\label{caap}
Let $l_1\rightarrow r_1$ and $l_2\rightarrow r_2$ be rules in $R$.
Let $\alpha\in POS(r_1)$, where $r_1/\alpha\in \ts(X_j)$, $j\geq 0$. 
Let $\beta\in POS(l_2)$,  where  $l_2/\beta \in \ts(X)-X$,
and let $s\in \bar{T}_\S(X_{k})-X$, $k\geq 1$, 
be a supertree of $l_2/\beta$.
Let $\alpha=\lambda$ or $\beta=\lambda$.
Let 
\begin{equation}\label{oil}
(r_1/\alpha)[e_1, \ldots, e_j]=s[ z_1, \ldots, z_k]\, ,
\end{equation}
where $e_1, \ldots , e_j\in  sub(L \cup E) $, 
$z_1, \ldots, z_k\in T_{\S}$.
Let $\gamma\in POS(s)$ be such that $l_2/\beta\gamma\in X$, and
$s/\gamma=x_\nu$, 
for some $1\leq \nu \leq k$. Then 
$z_\nu\in  sub(L \cup E) $.
\end{lem}
\pr Let $l_1\in \ts(X_m)$ for some $m\geq 0$.
Let $l_3=s[x_{m+1}, \ldots , x_{m+k}]$. Then
$l_3\in \ts(\{\, x_{m+1}, \ldots , x_{m+k}\,\})$  is a supertree
of $l_2/\beta$, 
for each $m+1 \leq i \leq m+k$, $x_i$ appears exactly once in $l_3$.
Moreover, 
$var(l_1)\cap var(l_3)=\emptyset$, 
and by (\ref{oil}), 
\begin{equation}\label{felgyo}
(r_1/\alpha)[e_1, \ldots, e_j]=
l_3[x_{m+1}\leftarrow z_1, \ldots , x_{m+k}\leftarrow z_k]\, .
\end{equation}
Let $\s_1:X\rightarrow \ts(X)$ be a most general unifier  of $r_1/\alpha$ and $l_3$.
By (\ref{felgyo}), there is a substitution 
$\s_2: X \rightarrow T_\S(X)$ such that 
$$\s_2(\s_1(r_1/\alpha))=
(r_1/\alpha)[e_1, \ldots, e_j]=l_3[x_{m+1}\leftarrow z_1, \ldots, 
x_{m+k}\leftarrow z_k]
=\s_2(\s_1(l_3))\, ,$$ where
$\s_2(\s_1(x_i))=e_i$ for $1\leq i \leq j$ and 
$\s_2(\s_1(x_{m+i}))=z_i$ for $1\leq i \leq k$.
Let $\gamma\in POS(s)$ be such that $l_2/\beta\gamma\in X$, and
$s/\gamma=x_\nu$, 
for some $1\leq \nu \leq k$.
By Definition \ref{GSM} and by the definition of $E$, $\s_1(x_{m+\nu})\in X\cup sub(E)$.
If $\s_1(x_{m+\nu})\in X$, then 
$\s_2(\s_1(x_{m+\nu}))$ is a subtree of $e_\mu$ for some 
$\mu\in  \{\, 1, \ldots, j\,\}$. Hence by the definition of $e_1, \ldots, e_j$, 
$z_{\nu}=\s_2(\s_1(x_{m+\nu}))\in  sub(L \cup E) $.
If $\s_1(x_{m+\nu})\in sub(E)$, then $z_{\nu}=\s_2(\s_1(x_{m+\nu}))
=\s_1(x_{m+\nu})\in sub(E)$.
\hfill $\b$

\vspace{2.5mm}
Intuitively, the following lemma states that along a reduction sequence of $S$
we   can reverse the order of the consecutive 
application of  
a  $S_{0}$-rule at $\alpha\in N^*$ and the application of an $(S-S_0)$-rule
at $\beta\in N^*$ if $\alpha$ is not a prefix of $\beta$ and $\beta$ is not a prefix of $\alpha$.
\begin{lem}\label{brics}
Let $$u_1\red S u_2 \red S u_3$$
 be a reduction sequence of $\cal C$, where $u_1, u_2, u_3 \in T_{\S \cup B}$. 
Let $\alpha\in POS(u_1)$, 
and $\beta\in POS(u_2)$ be such that 
$u_1\red S u_2$
applying a  rule ${\bf rule_1}$ of $S_0$ at $\alpha$, and that
$u_2\red S u_3$ applying an $(S-S_0)$-rule ${\bf rule_2}$ at $\beta$.
If $\alpha$ is not a prefix of $\beta$ and $\beta$ is not a prefix of $\alpha$,
then there is a tree $v\in T_{\S \cup B}$ such that 
$u_1\red S v$ applying ${\bf rule_2}$ at $\beta$, and
$v\red S u_3$ applying ${\bf rule_1}$ at $\alpha$.
\end{lem}
\pr Straightforward.
\hfill $\b$

\begin{lem} \label{twentyfive}
Let $i\geq 0$, 
$t\in \bar{T}_{\S\cup B}(X_1)$, $\alpha\in POS(t)$, $t/\alpha=x_1$,
$p \in D-sub(L)$, and 
$w\in sub(L)$.
Let 
\begin{equation}\label{kinai}
t[\langle p \rangle ]=u_1\red {S_i} u_2 \red {S_i}\ldots \red {S_i}u_n=\langle w \rangle 
\end{equation}
with $n\geq 1$, $u_1, \ldots , u_n\in T_{\S\cup B}$.
Then along {\em (\ref{kinai})}, 
${\cal C}_i$ applies a rule in $S_i-S_0$
at some prefix $\beta$ of $\alpha$.
\end{lem}
\pr By  direct inspection of the construction of the 
${\cal C}_i$'s.
\hfill $\b$

\begin{lem}\label{one}
For any $n \geq 0$, $u\in \bar{T}_\S(X_n)$,  
$v_1 , \ldots ,v_n,
 v \in D$, $m\geq 1$, 
and $w_1$, $\ldots$ , $w_m \in T_{\S\cup B}$,  if 
\begin{equation}\label{pioca}
u[\langle v_1 \rangle, \ldots ,\langle v_n \rangle]=
w_1 \red {S_0} w_2 \red {S_0}\ldots \red {S_0} w_m=
 \langle v \rangle,
\end{equation}
 then
$u[v_1, \ldots, v_n]=v$.
\end{lem}
\pr We proceed by induction on $height(u)$. The basis $height(u)=0$
of the induction  is trivial.
The induction step is a simple consequence of 
the definition of $S_0$.

\hfill $\b$


\begin{lem}\label{two}
Let $t\in L({\cal C})$, $m\geq 1$, 
$t_1, \ldots, t_m\in T_{\S\cup B}$, $ b \in A'$,
 and let 
\begin{equation}\label{keplet2}
t=t_1\red S t_2 \red S t_3 \red S \ldots \red S t_m=b\, .
\end{equation}
Let $l\rightarrow r$ be a rule in $R$, where $l\in
 \bar{T}_\S(X_n)$ and 
$n\geq 1$.
Moreover, let $1\leq j \leq m$, and let 
\begin{equation}\label{dino}
t_j/\alpha=l[\langle v_1 \rangle, \ldots ,\langle v_n \rangle]\,,
\end{equation}
where $n \geq 1$, 
$v_1 , \ldots ,v_n \in D$,
$\alpha \in POS(t_j)$. Let
$\alpha_1, \ldots ,\alpha_n\in POS(l)$ be 
such that
\begin{equation} \label{zsipo}
l/\alpha_i=x_i \mbox{ for }1\leq i \leq n\, .
\end{equation}
Consider the  reduction subsequence
\begin{equation}\label{keplet3}
t_j \red S t_{j+1} \red S \ldots \red S t_m=b
\end{equation} of {\em (\ref{keplet2})}.
If $\cal C$ does not apply any rules
at the 
positions $\alpha\alpha_1, \ldots ,\alpha\alpha_n$ along 
{\em (\ref{keplet3})}, then 
$v_1, \ldots, v_n\in  sub(L\cup E)$.
\end{lem}
\pr 
Let $1\leq i \leq n$, and let us assume that
$v_i \in D- sub(L)$. By (\ref{dino}) and (\ref{zsipo}), 
\begin{equation}\label{zorba}
t_j/\alpha\alpha_i=\langle v_i \rangle\, .
\end{equation}
 By Lemma \ref{twentyfive}, $\cal C$ applies a rule in 
$S-S_0$ at some prefix of $\alpha\alpha_i$
along (\ref{keplet3}).
Let $\beta\in POS(t_j)$  be the longest
prefix of $\alpha\alpha_i$ such that $\cal C$ 
applies a rule {\bf rule} in 
$S-S_0$ at $\beta$ along (\ref{keplet3}).
Then {\bf rule} is of the form  
$\langle r_1[e_1, \ldots , e_\kappa] \rangle \rightarrow c$, where 
$\kappa\geq 0$,
$r_1\in T_\S(X_\kappa)$, $e_1, \ldots, e_\kappa\in  sub(L \cup E) $, and
there is a rule 
$l_1\rightarrow r_1$ in $R$. Moreover 
there exists $\xi$, $j< \xi \leq m$, such that

$$
t_j/\beta \tred S 
t_{j+1}/\beta  \tred S \ldots \tred S t_\xi/\beta=
\langle r_1[e_1, \ldots, e_\kappa]\rangle \,,
$$
where for each $\pi$, $j\leq \pi \leq \xi -1$, 
$t_\pi/\beta=t_{\pi+1}/\beta$ or 
$t_\pi/\beta\red S t_{\pi+1}/\beta$.
We lose no generality by assuming that 
\begin{equation}\label{ubul}
t_j/\beta \red S 
t_{j+1}/\beta  \red S \ldots \red S t_\xi/\beta=
\langle r_1[e_1, \ldots, e_\kappa]\rangle \, .
\end{equation}
By Lemma \ref{brics} we  
may assume that there exists  $\nu $,
$j\leq \nu\leq \xi$  such 
that 

(a) along the reduction subsequence 
\begin{equation}\label{mohacs}
t_j/\beta\red S \ldots \red S t_\nu/\beta
\end{equation}
of (\ref{ubul})  no rule is applied at any prefix of $\alpha\alpha_i$, 
that

(b) along (\ref{mohacs})
each application of a rule of  $S_0$ at some $\delta\in N^*$
is followed somewhere later 
by an application of an $S-S_0$-rule of $S$ at a prefix
$\epsilon$ of $\delta$,
and that 

(c) along the reduction subsequence
$$
t_\nu/\beta\red S \ldots \red S t_\xi/\beta=
\langle r_1[e_1, \ldots, e_\kappa]\rangle 
$$
of (\ref{ubul}),
$S$ applies only rules of  $S_0$.

\noindent Then 
\begin{equation}\label{drei}
t_\nu/\beta=s[\langle z_1 \rangle, \ldots, \langle z_k
\rangle ]
\end{equation}
for some $k\geq 1$, $s\in \bar{T}_\S(X_k)$, and
$z_1, \ldots, z_k\in D$.
By (\ref{drei}), (c) of the definition of $\nu$, 
and Lemma \ref{one},
\begin{equation} \label{zwei}
s[z_1, \ldots, z_k]=r_1[e_1, \ldots, e_\kappa]\, .
\end{equation}
The word $\alpha$ is a prefix of $\beta$ or $\beta$ is  a prefix of 
 $\alpha$. Hence we can distinguish two cases.

\begin{figure}
\unitlength=1.00mm
\linethickness{0.4pt}
\begin{picture}(110.00,80.00)
\put(60.00,80.00){\line(-2,-3){50.00}}
\put(10.00,5.00){\line(1,0){100.00}}
\put(110.00,5.00){\line(-2,3){50.00}}
\put(60.00,50.00){\line(-4,-5){36.00}}
\put(60.00,50.00){\line(4,-5){36.00}}
\put(60.00,25.00){\line(-1,-1){20.00}}
\put(60.00,25.00){\line(1,-1){21.00}}
\bezier{132}(60.00,79.00)(53.00,59.00)(60.00,49.00)
\bezier{104}(60.00,49.00)(65.00,33.00)(60.00,25.00)
\bezier{92}(60.00,25.00)(55.00,11.00)(60.00,5.00)
\put(59.00,63.00){\makebox(0,0)[cc]{$\alpha$}}
\put(59.00,36.00){\makebox(0,0)[cc]{$\gamma$}}
\put(60.00,12.00){\makebox(0,0)[cc]{$\delta$}}
\put(59.00,1.00){\makebox(0,0)[cc]{$\langle v_i\rangle$}}
\put(50.00,75.00){\makebox(0,0)[cc]{$t_j$}}
\put(107.00,75.00){\makebox(0,0)[cc]{$t_j/\alpha=l[\langle v_1\rangle, 
\ldots , \langle v_n \rangle]$}}
\put(94.00,68.00){\makebox(0,0)[cc]{$\beta = \alpha\gamma$}}
\put(94.00,61.00){\makebox(0,0)[cc]{$\alpha_i = \gamma\delta$}}
\put(94.00,54.00){\makebox(0,0)[cc]{$\beta\delta=\alpha\alpha_i$}}
\end{picture}
\caption{Case 1.}
\end{figure}

{\bf Case 1}  $\alpha$ is a prefix of $\beta$, see Figure 4.
 In this case,  
\begin{equation}\label{vier}
\beta
=\alpha\gamma
\end{equation} 
for some $\gamma\in N^*$, and
 hence $t_\nu/\beta$ is a subtree of $t_\nu/\alpha$. 
Now by (\ref{dino}), the definition of $\nu$, and 
(\ref{drei}), 
\begin{equation}\label{radir}
s \mbox{ is a supertree of }l/\gamma\, .
\end{equation}

Let $\omega$ be the pefix of $\alpha\alpha_i$ with $length(\omega)=
length(\alpha\alpha_i)-1$. Observe that $\cal C$ applies a 
rule of  $S_0$ at the position $\omega$
along (\ref{keplet3}). Hence
\begin{equation}\label{vicki}
s\not\in X\, .
\end{equation}
We define  $\delta\in N^*$ be  by the equation $\gamma\delta=\alpha_i$.
Then
\begin{equation}\label{vasarh}
\beta\delta=\alpha\alpha_i\, ,
\end{equation} 
and by (a) of the definition of $\nu$, 
\begin{equation}\label{dune}
\delta\in POS(s), \hspace{1.5mm} \delta\in POS(l/\gamma), \mbox{ and }
(l/\gamma)/\delta=x_i\, . 
\end{equation}
By (\ref{vasarh}) and by (a) of the definition of $\nu$,
$$\beta\delta\in POS(t_\nu)\, .$$
By (\ref{zorba}), (\ref{vasarh}), (a) of the definition of $\nu$, 
and (\ref{drei}),
\begin{equation}\label{athen}
\langle v_i\rangle=(t_j/\beta)/\delta=(t_\nu/\beta)/\delta=
s[\langle z_1 \rangle, \ldots , \langle z_k \rangle ]/\delta=\langle z_\mu
\rangle
\end{equation}
for some $ 1\leq\mu \leq k$.
As $R$ is a GSM-TRS, 
by  (\ref{radir}), (\ref{vicki}), (\ref{dune}), (\ref{zwei}),
and  Lemma \ref{caap}, $z_\mu\in  sub(L \cup E) $.
By (\ref{athen}), $v_i=z_\mu$. Thus $v_i\in  sub(L \cup E) $.

\begin{figure}
\unitlength=1.00mm
\linethickness{0.4pt}
\begin{picture}(110.00,80.00)
\put(60.00,80.00){\line(-2,-3){50.00}}
\put(10.00,5.00){\line(1,0){100.00}}
\put(110.00,5.00){\line(-2,3){50.00}}
\put(60.00,50.00){\line(-4,-5){36.00}}
\put(60.00,50.00){\line(4,-5){36.00}}
\put(60.00,25.00){\line(-1,-1){20.00}}
\put(60.00,25.00){\line(1,-1){21.00}}
\bezier{132}(60.00,79.00)(53.00,59.00)(60.00,49.00)
\bezier{104}(60.00,49.00)(65.00,33.00)(60.00,25.00)
\bezier{92}(60.00,25.00)(55.00,11.00)(60.00,5.00)
\put(59.00,63.00){\makebox(0,0)[cc]{$\beta$}}
\put(59.00,36.00){\makebox(0,0)[cc]{$\gamma$}}
\put(60.00,12.00){\makebox(0,0)[cc]{$\alpha_i$}}
\put(59.00,1.00){\makebox(0,0)[cc]{$\langle v_i\rangle$}}
\put(50.00,75.00){\makebox(0,0)[cc]{$t_j$}}
\put(89.00,65.00){\makebox(0,0)[cc]{$\alpha =\beta\gamma$}}
\put(102.00,75.00){\makebox(0,0)[cc]{$t_j/\alpha=l[\langle v_1\rangle, 
\ldots , \langle v_n \rangle]$}}
\end{picture}
\caption{Case 2.}
\end{figure}

\vspace{2.5 mm}
{\bf Case 2} $\beta$ is a prefix of $\alpha$, see Figure 5. In this case
\begin{equation}\label{seven}
\alpha=\beta\gamma
\end{equation} 
for some $\gamma\in N^*$, and
 hence $t_j/\alpha$ is a subtree of $t_j/\beta$. 
Now by  (\ref{dino}), the definition of $\nu$, and 
(\ref{drei}), 
\begin{equation}\label{blazsikz}
s/\gamma \mbox{ is a supertree of } l\, .
\end{equation}
Moreover, by (a) of the definition of $\nu$,
\begin{equation}\label{cocacola}
\alpha_i\in POS(s/\gamma), \hspace{1.5mm} l/\alpha_i\in X, 
\mbox{ and }(s/\gamma)/\alpha_i\in X\, .
\end{equation}
Let $\omega$ be the pefix of $\alpha\alpha_i$ with $length(\omega)=
length(\alpha\alpha_i)-1$. Observe that $\cal C$ applies a 
rule of  $S_0$ at the position $\omega$
along (\ref{keplet3}). Hence
\begin{equation}\label{ajax}
s/\gamma\not\in X\, .
\end{equation}
By (\ref{seven}) and by (a) of the definition of $\nu$, 
\begin{equation}\label{korfu}
\beta\gamma\alpha_i=\alpha\alpha_i\in POS(t_\nu)\, .
\end{equation}
Then by (\ref{zorba}), (\ref{korfu}), 
(a) of the definition of $\nu$, and (\ref{drei}), 
\begin{equation}\label{athen2}
\langle v_i\rangle=(t_j/\beta)/\gamma\alpha_i=
(t_\nu/\beta)/\gamma\alpha_i=
s[\langle z_1 \rangle, \ldots , \langle z_k \rangle ]/\gamma\alpha_i
=\langle z_\mu\rangle
\end{equation}
for some $1\leq\mu \leq k$.
By (\ref{zwei}), 
\begin{equation}\label{nielsen}
(s/\gamma)[z_1, \ldots, z_k]=
s[z_1, \ldots, z_k]/\gamma=r_1[e_1, \ldots, e_\kappa]/\gamma\, .
\end{equation}
As $R$ is a GSM-TRS, by (\ref{blazsikz}), (\ref{ajax}),
(\ref{cocacola}), (\ref{athen2}),
(\ref{nielsen}), and Lemma \ref{caap}, $z_\mu\in  sub(L \cup E) $.
By (\ref{athen2}), $v_i=z_\mu$. Thus $v_i\in  sub(L \cup E) $. 
\hfill $\b$

\begin{lem} \label{masodik} 
$R^*(L) \subseteq L({\cal C})$.
\end{lem}
\pr By (\ref{martzi}), 
 $L\subseteq L({\cal C}_0)$. As $S_{i-1}\subseteq S_i$ for $i\geq 1$, 
we have $L\subseteq L({\cal C}_i)$ for $i\geq 0$. Hence $L\subseteq 
L({\cal C})$. Thus it is sufficient to show that for each $t\in L({\cal C})$, 
if $t\red R t'$, then $t'\in L({\cal C})$.
To this end, let us suppose that $t\red R t'$, applying 
the rule $l\rightarrow r$ in $R$ at $\alpha\in POS(t)$. 
Here $l\in \bar{T}_\S(X_n)$ for
some $n\geq 0$. Let $\alpha_1, \ldots , \alpha_n\in POS(l)$ be such
that 
$$
l/\alpha_i=x_i \mbox{ for } 1\leq i \leq n\, .
$$
Then 
$$t=s[l[u_1, \ldots , u_n]]\, ,$$
where $s\in \bar{T}_\S(X_1)$, $\alpha\in POS(s)$, $s/\alpha=x_1$, and
$u_1, \ldots, u_n\in \ts$. 
Moreover, 
$$t'=t[\alpha\leftarrow r[u_1, \ldots , u_n]]=s[r[u_1, \ldots , u_n]]\, .$$
As $t\in L({\cal C})$, there is a reduction 
sequence 
\begin{equation}\label{dollar}
t=t_1\red S t_2 \red S t_3 \red S \ldots \red S t_m =b,
\end{equation}
where $m\geq 1$, $b\in A'$, $t_1, \ldots , t_m\in T_{\S\cup B}$, and there
are integers $j, k$ with $1\leq j \leq k \leq m$ such that

(i) $t_j=s[l[\langle v_1 \rangle, \ldots , \langle v_n \rangle]]$, where
$v_i\in D$ and 
$u_i \tred S \langle v_i \rangle$  for $1\leq i \leq n$, 

(ii) $t_k=s[c_0]$, for some $c_0\in A$, 
where $l[\langle v_1\rangle, \ldots , \langle v_n \rangle ]\tred
S c_0$, 
and that

(iii) along the reduction subsequence  $t_j \red S t_{j+1} \red S \ldots
\red S t_k$ of (\ref{dollar}), $\cal C$ does not apply any  rules
at the positions $\alpha\alpha_1, \ldots ,
\alpha\alpha_n$.
By Lemma \ref{two}, $v_1, \ldots, v_n\in  sub(L \cup E) $.
Hence by Condition (b) in the definition of $S_i$, $i\geq 1$, 
and by the definition of $\cal C$, the rule 
$r[\langle v_1\rangle , \ldots , \langle v_n\rangle ]
\rightarrow c_0$ is in $S$.
Thus we get 
$$t'=s[r[u_1, \ldots , u_n]]\tred S
s[r[\langle v_1 \rangle, 
\ldots , \langle v_n \rangle ]]\red S s[c_0]\tred S b\, .$$
As $b\in A'$, we have $t'\in L({\cal C})$.
\hfill $\b$

\vspace{2.5 mm}
By Lemma  \ref{elso} and Lemma \ref{masodik}, we get that
$R^*(L)= L({\cal C})$. 

\hfill $\b$

Lemma  \ref{pamir} and Theorem \ref{fotetel} imply the following result. 
\begin{tet} There is a left-linear monadic TRS $R$ over a ranked alphabet 
$\S$ such that  $R$ is an  EPRF-TRS and that $R$ is not a P$\S$R-TRS. 
\end{tet}
\section{An Example}\label{negyedik}
We illustrate the construction of 
${\cal C}_j$, $j\geq 0$, appearing in the previous section 
by an  example.
Let $\S=\S_0\cup \S_1 \cup\S_3$, $\S_0=\{\, \sharp\,\}$, 
$\S_1=\{\, f \,\}$, $\S_3=\{\, g\,\}$. Let the  TRS $R$
over $\S$  consist of the following two rules.

$$f(f(g(x_1, \sharp, \sharp)))\rightarrow f(f(x_1))\, ,$$
$$g(x_1, x_2, \sharp)\rightarrow f(g(x_1, \sharp, x_1))\, .$$
By direct inspection we obtain that $R$ is a left-linear GSM-TRS.
Here $E=\{\, \sharp \,\}$.
Let $L=\{\, g(\sharp,\sharp, \sharp)\,\}$.
Then $sub(L\cup E)=\{\,\sharp,  g(\sharp,\sharp, \sharp)\,\}$.
It is not hard to see that 
$$R^*(L)=\{\, f^n(g(\sharp, \sharp,\sharp))\mid n\geq 0\,\}\cup 
\{\, f^n(\sharp)\mid n\geq 2\,\}\, .$$
By direct inspection we obtain that the set of subterms of the right-hand 
sides of the rules of $R$ is 
$$\{\, x_1, f(x_1), f(f(x_1)), \sharp, g(x_1, \sharp,x_1), f(g(x_1, \sharp,x_1))
\,\}\, .$$
Then
 
$D=
\{\, \sharp, \, f(\sharp), 
\, g(\sharp,\sharp, \sharp),  \, f(f(\sharp)),    \, f(g(\sharp,\sharp, \sharp)), \, 
  \, f(f(g(\sharp,\sharp, \sharp))),   \, $
 $g(g(\sharp,\sharp, \sharp), \sharp, g(\sharp,\sharp, \sharp))$,

$
f(g(g(\sharp,\sharp, \sharp), \sharp, g(\sharp,\sharp, \sharp)))
\,\}\, . $

 ${\cal C}_0=(\S, B, S_0, \{\, \langle g(\sharp, \sharp, \sharp)\rangle\, \})$, 
where 

$B=\{\, \langle \sharp \rangle,  \, \langle g(\sharp,\sharp, \sharp) \rangle,  \, 
\langle f(\sharp) \rangle,    \, \langle f(g(\sharp,\sharp, \sharp)) \rangle, \, 
\langle f(f(\sharp)) \rangle,  \, \langle f(f(g(\sharp,\sharp, \sharp))) \rangle,  \, $ 
 
$\langle g(g(\sharp,\sharp, \sharp), \sharp, g(\sharp,\sharp, \sharp)) \rangle, 
  \, \langle f(g(g(\sharp,\sharp, \sharp), \sharp, g(\sharp,\sharp, \sharp))) \rangle
\,\}.$
 
\noindent
Furthermore, 
$S_0$ consists of the following eight rules.

$\sharp \rightarrow \langle \sharp \rangle$, 

$g(\langle\sharp\rangle, \langle \sharp\rangle, \langle \sharp\rangle)\rightarrow 
\langle g(\sharp,\sharp, \sharp)\rangle$, 
 
$f(\langle\sharp\rangle)\rightarrow 
\langle f(\sharp)\rangle$,

$f(\langle g(\sharp,\sharp, \sharp)\rangle) \rightarrow 
\langle f(g(\sharp,\sharp, \sharp))\rangle$,

$f(f\langle(\sharp)\rangle)\rightarrow  \langle f(f(\sharp))\rangle$,

$f(\langle f(g(\sharp,\sharp, \sharp))\rangle)\rightarrow \langle  f(f(g(\sharp,\sharp, \sharp)))\rangle$,

 $g(\langle g(\sharp,\sharp, \sharp)\rangle, \langle \sharp\rangle, \langle g(\sharp,\sharp, \sharp)\rangle)
\rightarrow \langle g(g(\sharp,\sharp, \sharp), \sharp, g(\sharp,\sharp, \sharp))\rangle$,

$
f(\langle g(g(\sharp,\sharp, \sharp), \sharp, g(\sharp,\sharp, \sharp))\rangle )\rightarrow 
\langle f(g(g(\sharp,\sharp, \sharp), \sharp, g(\sharp,\sharp, \sharp)))\rangle $.

${\cal C}_1=(\S, B, S_1, \{\, \langle   g(\sharp,\sharp,\sharp )\rangle)\, \}$, where 
$S_1$ contains all rules of $S_0$ and the following   rules.

$\langle f(f(\sharp))\rangle \rightarrow  \langle f(f(g(\sharp,\sharp, \sharp))) \rangle ,$

$\langle f(g(\sharp,\sharp, \sharp))\rangle \rightarrow \langle g(\sharp,\sharp,\sharp )
\rangle ,$

 ${\cal C}_2=(\S, B, S_2, \{\, \langle   g(\sharp,\sharp,\sharp )\rangle\, \})$, where 
$S_2$ contains all rules of $S_1$ and the following two rules.

$\langle f(f(\sharp))\rangle \rightarrow \langle f(g(\sharp,\sharp,\sharp ))
\rangle ,$

$\langle f(f(\sharp))\rangle \rightarrow \langle g(\sharp,\sharp,\sharp )
\rangle .$


The bta 
 ${\cal C}_3=(\S, B, S_3, \{\, \langle   g(\sharp,\sharp,\sharp )\rangle\, \})$ is equal to ${\cal C}_2$. 
By direct inspection we obtain that the states  

$ \langle f(f(g(\sharp,\sharp, \sharp))) \rangle,\,  
\langle g(g(\sharp,\sharp, \sharp), \sharp, g(\sharp,\sharp, \sharp)) \rangle$,  

$ \langle f(g(g(\sharp,\sharp, \sharp), \sharp, g(\sharp,\sharp, \sharp))) \rangle$

\noindent 
 are superfluous as  the final state $\langle g(\sharp, \sharp, \sharp)\rangle$ 
cannot be reached from any of them. Hence we drop all of them and also 
omit all rules in which they appear.
In this way we obtain the bta 
${\cal B}_1=(\S, B_1, Q_1, \{\, \langle  g(\sharp,\sharp,\sharp ) \rangle \, \})$, where

$B_1=\{\, \langle \sharp \rangle,  \, \langle g(\sharp,\sharp, \sharp) \rangle,  \, 
\langle f(\sharp) \rangle,    \, \langle f(g(\sharp,\sharp, \sharp)) \rangle, \, 
\langle f(f(\sharp)) \rangle\, \}$ and  

$ Q_1$ consists of the following 
 rules.

$\sharp \rightarrow \langle \sharp \rangle$,

$g(\langle\sharp\rangle, \langle \sharp\rangle, \langle \sharp\rangle)\rightarrow 
\langle g(\sharp,\sharp, \sharp)\rangle$, 
 
$f(\langle\sharp\rangle)\rightarrow 
\langle f(\sharp)\rangle$,

$f(f\langle(\sharp)\rangle)\rightarrow  \langle f(f(\sharp))\rangle$,

$\langle f(g(\sharp,\sharp, \sharp))\rangle \rightarrow \langle g(\sharp,\sharp,\sharp )
\rangle ,$

$\langle f(f(\sharp))\rangle \rightarrow \langle f(g(\sharp,\sharp,\sharp ))
\rangle ,$


$\langle f(f(\sharp))\rangle \rightarrow \langle g(\sharp,\sharp, \sharp) \rangle,$

\noindent 
We obtain the bta 
${\cal B}_2=(\S, B_2, Q_2, \langle   \{\, g(\sharp,\sharp,\sharp )\rangle\, \})$ from ${\cal B}_1$ by eliminating the lambda rules.
Here $B_2=B_1$ and $Q_2$ consists of the following 
rules.

$\sharp \rightarrow \langle \sharp \rangle$,

$g(\langle\sharp\rangle, \langle \sharp\rangle, \langle \sharp\rangle)\rightarrow 
\langle g(\sharp,\sharp, \sharp)\rangle $, 
 
$f(\langle\sharp\rangle)\rightarrow 
\langle f(\sharp)\rangle $,

$ f(\langle f(\sharp) \rangle ) \rightarrow  \langle f(f(\sharp)) \rangle$. 

$f(\langle f(\sharp))\rangle )\rightarrow \langle f(g(\sharp,\sharp,\sharp ))
\rangle ,$

$f(f\langle(\sharp)\rangle)\rightarrow  \langle g(\sharp,\sharp,\sharp )\rangle$,

$f(\langle g(\sharp,\sharp, \sharp) \rangle ) \rightarrow \langle g(\sharp,\sharp,\sharp )\rangle $. 

By direct inspection we obtain that the states  
states 
$\langle f(f(\sharp)) \rangle$ and $\langle f(g(\sharp,\sharp, \sharp)) \rangle$

\noindent 
 are superfluous as  the final state $\langle g(\sharp, \sharp, \sharp)\rangle$ 
cannot be reached from any of them. Hence we drop all of them and also 
omit all rules in which they appear.

\noindent 
In this way we obtain the bta 
${\cal B}_3=(\S, B_3, Q_3, \{\, \langle   g(\sharp,\sharp,\sharp )\rangle\, \})$.

Here 
$B_3=\{\, \langle \sharp \rangle,  \, \langle g(\sharp,\sharp, \sharp) \rangle,  \, 
\langle f(\sharp) \rangle,    \,  \}$ and

$Q_3$ consists of the following five
rules.

$\sharp \rightarrow \langle \sharp \rangle$,

$g(\langle\sharp\rangle, \langle \sharp\rangle, \langle \sharp\rangle)\rightarrow 
\langle g(\sharp,\sharp, \sharp)\rangle $, 
 
$f(\langle\sharp\rangle)\rightarrow 
\langle f(\sharp)\rangle $,

$f(\langle f(\sharp)\rangle)\rightarrow  \langle g(\sharp,\sharp,\sharp )\rangle$,

$f(\langle g(\sharp,\sharp, \sharp) \rangle ) \rightarrow \langle g(\sharp,\sharp,\sharp )\rangle $.

\noindent Then $L( {\cal C}_3)  =L({\cal B}_3)$.
We obtain by direct inspection that $L({\cal B}_3)=R^*(L)$.

\section{PRF-TRSs}\label{marci}

We show various decidability and undecidability results on 
PRF-TRSs and EPRF-TRSs. We show that
 reachability, joinability, and  local confluence are  decidable for   EPRF-TRSs.


\begin{tet} There is a ranked alphabet $\S$ and there is a linear  EP$\S$RF-TRS
$R$ such that $R$ is not a PRF-TRS.
\end{tet}
\pr Let $\S=\S_1 \cup \S_0$,  $\S_1=\{\, f, g\, \}$, 
$\S_0=\{\, \sharp \, \}$. Let $R$ consist of the following five
rules.

$f(g(x_1)\rightarrow f(f(g(g(x_1)))),$ 

 $f(\sharp)\rightarrow \sharp,$

$g(\sharp)\rightarrow \sharp,$

$\sharp \rightarrow f(\sharp),$ 

$\sharp \rightarrow g(\sharp).$

\noindent It should be clear that for each tree $t\in \ts$, $t\tred R \sharp$
and  $\sharp \tred R t$. Hence for each nonempty tree
language $L\subseteq \ts$,
$R^*(L)=\ts$. Thus $R$ 
is an EP$\S$RF-TRS. 

Let $\Delta=\S \cup \{\, h\, \}$, where $h\in \Delta_1$.
 Then 
 $R^*( \{\, f(g(h(\sharp)))\, \}   )=\{\, f^n(g^n(h(t)))\mid n\geq 0, t\in \ts\,\}$ 
 is not recognizable.
\hfill $\b$


\begin{tet}\label{cucor}
Let $R$ be any TRS over $sign(R)$, and let 
$\S=\{\, f, \sharp \, \}\cup sign(R)$, where 
 $f\in \S_2-sign(R)$ and $\sharp\in \S_0-sign(R)$. Then 
$R$ is a P$\S$RF-TRS if and only if $R$ is a PRF-TRS.
\end{tet}
\pr ($\Leftarrow$) Trivial.

($\Rightarrow$) Let $\Gamma$ be an arbitrary ranked alphabet with 
$sign(R)\subseteq \Gamma$. 
To each symbol $g\in \Gamma_k-sign(R)$, $k\geq 0$, we assign a tree 
$t_g\in \ts(X_k)$. To this end,
we number the symbols in $\Gamma-sign(R)$
from $1$ to $|\Gamma-sign(R)|$.

Then  we define the $n$th right comb $right_n$ for $n\geq 0$,  as follows.

(i)
$right_0=\sharp$,

(ii)
for each $n\geq 0$, 
$right_{n+1}=f(\sharp, right_n)$.

\noindent
For the definition of  the $n$th left comb
$left_n$, see the proof of Statement \ref{pamir}. 
To any  symbol $g\in \Gamma_k-sign(R)$, $k\geq 0$, with number
$l$, 
we assign the  tree $t_g=f(left_k, right_l)$.

Consider the TRS 
$$S=\{\, g(x_1, \ldots, x_k)\rightarrow t_g
\mid k\geq 0, g\in \Gamma_k-sign(R)\,\}\, .$$
It should be clear that $S$ is a convergent TRS. 
It is not hard to show the following two statements.
\begin{cla}\label{claim1}
For any $r,s\in T_\Gamma$, 
$$r\red R s \mbox{ if and only if } r\hspace{-1.5mm} \downarrow_S\red R s\hspace{-1.5mm} \downarrow_S  \, .$$
\end{cla}

\begin{cla}\label{claim2}
A tree language $L$
over $sign(R)$ is finite  if and only if  the tree language $S(L)$ 
over $\S$  finite.
A tree language $L$
over $sign(R)$ is recognizable   if and only if  the tree language $S(L)$ 
over $\S$  recognizable.
\end{cla} 

Let $L$ be any finite  tree language over  $sign(R)$.
By Claim \ref{claim2}, $S(L)$ is a finite  tree language over $\S$.
By Claim \ref{claim1}, $S(R^*_{sign(R)}(L))=R^*_\S(S(L))$. 
By Claim \ref{claim2}, $R^*_{sign(R)}(L)$ is recognizable if and only if
$R^*_\S(S(L))$ is recognizable.
Hence if $R$ is a P$\S$RF-TRS, then $R$  is a P$\Gamma$RF-TRS.
As $\Gamma$ is  an arbitrary ranked alphabet with 
$sign(R)\subseteq \Gamma$, $R$ is a PRF-TRS.

\hfill $\b$

The proof of the following result is similar to the proof of Theorem
\ref{cucor}.

\begin{tet}
Let $R$ be a TRS over $sign(R)$, and let 
$\S=\{\, f, \sharp \, \}\cup sign(R)$, where 
 $f\in \S_2-sign(R)$ and $\sharp\in \S_0-sign(R)$. 
$R$ is an EP$\S$RF-TRS if and only if $R$ is an EPRF-TRS.
\end{tet}

\begin{cons}
Let $R$ be a TRS over $\S$ such that there is a symbol
$f\in \S_2-sign(R)$ and there is a constant 
$\sharp\in \S_0-sign(R)$. Then $R$ is a PRF-TRS 
if and only if $R$  $R$ is a P$\S$RF-TRS. Moreover,
$R$ is an EPRF-TRS if and only if $R$  is an EP$\S$RF-TRS.
\end{cons}
We now show that reachability is decidable for   EPRF-TRSs.
\begin{tet}\label{zenebona}
Let $R$ be an EPRF-TRS over $\S$  and let
$p, q\in \ts(X)$. Then it is decidable   whether   $p\tred R q$.
\end{tet}
\pr  
Let $m\geq 0$ be such that $var(p)\subseteq X_m$, $var(q)\subseteq X_m$.
Let  us introduce new constant symbols $Z=\{\, z_1, \ldots, z_m\,\}$ with
$Z\cap \S=\emptyset$.
For each $t\in \ts(X_m)$, we define $t_z\in T_{\S \cup Z}$ as 
$t_z=t[z_1, \ldots, z_m]$. 
By direct inspection we obtain that for all $u, v\in \ts(X)$, 
$$u \red R v \mbox{ if and only if } u_z \red R v_z \, , $$
hence
$$u \tred R v \mbox{ if and only if } u_z \tred R v_z  \, .$$
Consider the singleton set 
$\{\, p_z\,\}$.
As $R$ is an EPRF-TRS,
 $R^*_{\S\cup Z}(\{\, p_z\,\})$ is a  recognizable tree language over $\S\cup Z$,
 and we can construct a bta 
 over $\S\cup Z$
which recognize $R^*_{\S\cup Z}(\{\, p_z\,\})$.
Hence we can decide   whether
$ q_z\in R^*_{\S\cup Z}(\{\, p_z\, \})$, see \cite{gecste}.
Clearly, $ q_z\in R^*_{\S\cup Z}(\{\, p_z\, \})$
if and only if $p\tred R q$.

\hfill $\b$

We now show that  joinability is decidable for   EPRF-TRSs.
\begin{tet}\label{fu}
Let $R$ be an EPRF-TRS over $\S$, and let
$p, q\in \ts(X)$. Then it is decidable   whether   there is a tree $r\in \ts(X)$ such that
 $p\tred R r$ and $q\tred R r$. 
\end{tet}
\pr 
For each $t\in \ts(X_m)$, we define $t_z\in T_{\S \cup Z}$ as in the proof of Theorem \ref{zenebona}. 

\begin{cla}\label{zuhatag} For any $p, q\in \ts(X)$,  there is a tree $r\in \ts(X)$ such that
 $p\tred R r$ and $q\tred R r$  if and only if 
$R^*_{\S\cup Z}(\{\, p_z\,\})\cap R^*_{\S\cup Z}(\{\, q_z\,\})=\emptyset$.
\end{cla}
\pr 
Assume that $R^*_{\S\cup Z}(\{\, p_z\,\})\cap R^*_{\S\cup Z}(\{\, q_z\,\})=\emptyset$.
Then there is a tree $s\in R^*_{\S\cup Z}(\{\, p_z\,\})\cap R^*_{\S\cup Z}(\{\, q_z\,\})$. 
We define $r$ from $s$ by replacing each occurrence of $z_i$ by $x_i$ for $1\leq i \leq m$. 
Then  $p\tred R r$ and $q\tred R r$. 

Assume that  there is a tree $r  \ts(X)$ such that 
 $p\tred R r$ and $q\red r$. Hence $ r_z\in R^*_{\S\cup Z}(\{\, p_z\, \})$
and  $ r_z\in R^*_{\S\cup Z}(\{\, q_z\, \})$.
Thus $R^*_{\S\cup Z}(\{\, p_z\,\})\cap R^*_{\S\cup Z}(\{\, q_z\,\})=\emptyset$.

\hfill $\b$

As $R$  is an EPRF-TRS,
 $R^*_{\S\cup Z}(\{\, p_z\,\})$ and 
$R^*_{\S\cup Z}(\{\, q_z\,\})$ are  recognizable, and we can construct two btas
 over $\S\cup Z$
which recognize $R^*_{\S\cup Z}(\{\, p_z\,\})$ and $R^*_{\S\cup Z}(\{\, q_z\,\})$, respectively.
Hence we can decide   whether
$R^*_{\S\cup Z}(\{\, p_z\, \}) \cap R^*_{\S\cup Z} (\{\, q_z\, \})=
\emptyset$, see \cite{gecste}.
By Claim \ref{zuhatag}, if 
$R^*_{\S\cup Z}(\{\, p_z\,\})\cap R^*_{\S\cup Z}(\{\, q_z\,\})\neq \emptyset$, 
then there is a tree $r\in \ts(X)$ such that   $p\tred R r$ and $q\tred R r$. 
Otherwise, 
there is no tree $r\in \ts(X)$ such that   $p\tred R r$ and $q\tred R r$.

\hfill $\b$


\begin{tet}\label{muzsika}
Let $R$ be a confluent EPRF-TRS over $\S$, and let
$p, q\in \ts(X)$. Then it is decidable   whether   $p\tthue R q$.
\end{tet}
\pr $p\tthue R q$ if and only if there is a tree $r  \in \ts(X)$
such that  $p\tred R r$ and $q\red R r$. 
By Theorem \ref{fu}, we can decide whether 
there is a tree $r \in  \ts(X)$
such that  $p\tred R r$ and $q\tred Rr$.

\hfill $\b$


We now show that local confluence is decidable for   EPRF-TRSs.

\begin{tet} Let $R$ be an EPRF-TRS over $\S$.
 Then it is decidable  whether $R$ is locally confluent.
\end{tet}
\pr  It is well known  that 
$R$ is locally confluent
if and only if for every critical pair $(v_1, v_2)$ of $R$ there exists a 
tree $v\in \ts(X)$ such that $v_1\tred R v$ and $v_2\tred R v$, see \cite{baanip}.
Furthermore,  all critical pairs of $R$ are variants of 
finitely many critical pairs of $R$.
Hence  it is sufficient  to inspect finitely many critical pairs.
Thus the theorem follows from Theorem \ref{fu}. 
\hfill $\b$

\begin{tet} Let $R$ be an EPRF-TRS and $S$ be a TRS over $\S$.
 Then it is decidable whether $\tred {S}\subseteq \tred {R}$.
\end{tet}
\pr 
Let $m\geq 0$ be such that for all variables $x_i$ occurring on the left-hand 
side of some rule in $S$, $x_i\in X_m$, that is, $i\leq m$. From now on, 
for each $t\in \ts(X_m)$, we define $t_z\in T_{\S \cup Z}$ as in the proof of Theorem \ref{zenebona}.

\begin{cla} \label{tamas}
$\tred {S} \subseteq \tred {R}$
if and only if
for each rule $l\rightarrow r$ in $S$, 
$r_z\in R^*_{\S\cup Z}(\{\, l_z\, \})$. 
\end{cla}
\pr
$(\Rightarrow)$
Let $l\rightarrow r$ be  an arbitrary rule in $S$. Clearly, 
$l\tred {R} r$. Thus $r_z\in R^*_{\S\cup Z}(\{\, l_z\, \})$.

$(\Leftarrow)$ Let us suppose that $t_1, t_2 \in \ts(X)$, and that 
$t_1\red {S} t_2$ applying the rule $l\rightarrow r$. As
$r_z\in R^*_{\S\cup Z} (\{\, l_z\,\})$,
 $l_z\tred {R} r_z$ holds. Hence $l\tred {R} r$ implying that 
$t_1\tred {R} t_2$ 
as well.

\hfill $\b$ 

\vspace{2.5 mm}
Let  $l\rightarrow r$ be an arbitrary rule
in $S$. We can construct a bta 
over $\S\cup Z$ recognizing the singleton set 
$\{\, l_z\,\}$. As  $R$ is an EPRF-TRS, 
 $R^*_{\S\cup Z}(\{\, l_z\,\})$ is recognizable,
and we can construct a bta over $\S\cup Z$ recognizing
$R^*_{\S\cup Z}(\{\, l_z\,\})$.
Hence we can decide whether $r_z\in R^*_{\S\cup Z}(\{\, l_z\, \})$.
Thus by Claim \ref{tamas}, we can decide   whether
$\tred {S}\subseteq \tred {R}$.
\hfill $\b$

\begin{cons}\label{anna}
Let $R$ and $S$  be EPRF-TRS over $\S$. 
Then it is decidable which 
one of the following four mutually excluding conditions holds.

(i) $\tred {R}\subset  \tred {S}$,

(ii) $\tred {S}\subset  \tred {R}$,

(iii) $\tred {R}=  \tred {S}$,

(iv) $\tred {R}\Join \tred {S}$,

\noindent where ``$\Join$\,'' stands for the incomparability relationship.
\end{cons}

\begin{obs}\label{pollak}
If one omits a rule from a left-linear GSM-TRS, then the 
resulting rewrite system still remains a left-linear GSM-TRS.
\end{obs}

One can easily show  the following result applying Theorem \ref{fotetel},
Consequence \ref{anna}, and Observation \ref{pollak}.

\begin{cons}\label{evi}
For a left-linear GSM-TRS $R$, it is decidable whether 
$R$ is left-to-right minimal.
\end{cons}


Consequence \ref{anna} also implies the following.
\begin{cons}\label {eva}
Let $R$ and $S$  be  TRSs  over $\S$ such that $R\cup R^{-1}$ and 
$S\cup S^{-1}$ are EPRF-TRSs. 
Then it is decidable which 
one of the following four mutually excluding conditions holds.

(i) $\tthue {R}\subset  \tthue {S}$,

(ii) $\tthue {S}\subset  \tthue {R}$,

(iii) $\tthue {R}=  \tthue {S}$,

(iv) $\tthue {R}\Join \tthue {S}$.

\end{cons}


\begin{tet} Let $R$  be  an EPRF-TRS  and  $S$ be a TRS over a ranked alphabet $\S$.
Let 
$g\in \S-(sign(R)\cup \S_0)$. Let $\sharp \in \S_0$ be irreducible for $R$. 
 Then it is decidable whether  $\tred {S}\cap (\ts\times \ts)
\subseteq \tred {R}
\cap (\ts\times \ts)$.
\end{tet}
\pr 
We assume that $g\in \S_1$.
One can easily modify  the proof of this case when proving 
the more general case  $g\in \S_k$, $k\geq 1$.
For each $t\in \ts(X)$, we define $t_g \in T_{\S}$ from $t$ 
by substituting
$g^i(\sharp)$ for all occurrences of the variable  $x_i$ for $i\geq 1$.

\begin{cla} \label{tomi}
$\tred {S} \cap (\ts\times \ts)\subseteq \tred {R}\cap (\ts\times \ts)$
if and only
 if 
for each rule $l\rightarrow r$ in $S$, 
$r_g\in R^*_1(\{\, l_g\,\})$. 
\end{cla}
\pr 
$(\Rightarrow)$
Let $l\rightarrow r$ be  an arbitrary rule in $S$. Clearly, 
$l_g\red {S} r_g$. 
Thus by our assumption $l_g\tred {R} r_g$.

$(\Leftarrow)$ Let us suppose that $t_1, t_2 \in \ts$, and that 
$t_1\red {S} t_2$ applying the rule $l\rightarrow r$. As
$r_g\in R^*_1 (\{\, l_g\,\})$, 
$l_g\tred {R} r_g$ holds. Hence $l\tred {R} r$ 
implying that $t_1\tred {R} t_2$ 
as well.
\hfill $\b$ 

\vspace{2.5 mm}
\noindent
For each rule $l\rightarrow r$ in $S$, the tree language $\{\, l_g\,\}$
is recognizable, and we can construct a bta over $\S$ recognizing 
 $\{\, l_g\,\}$. As $R$ is an EPRF-TRS, 
$R^*(\{\, l_g\,\})$ is also recognizable, and we can construct a bta over $\S$
recognizing $R^*(\{\, l_g\,\})$. Hence for each rule  $l\rightarrow r$ in $S$,
we can decide whether or not $r_g\in R^*(\{\, l_g\,\})$.
Thus by Claim \ref{tomi}, we can decide whether  
$\tred {S} \cap (\ts\times \ts)\subseteq \tred {R}\cap (\ts\times \ts)$.
 \hfill $\b$


\begin{cons}\label{gabor}
Let $R$ and $S$  be EPRF-TRSs over $\S$.  Moreover,
let $g_1, g_2\in \S-\S_0$ be such that for each $i\in \{\, 1, 2\,\}$, 
$g_i$ does 
not occur on the
left-hand side of any rule in $R_i$. 
Let $\sharp_1, \sharp_2\in \S_0$ be  such that for each 
$i\in \{\, 1, 2\,\}$, $\sharp_i$ is irreducible for $R_i$.
Then it is decidable which 
one of the following four mutually excluding conditions holds.

(i) $\tred {R}\cap (\ts \times \ts )\subset  \tred {S}\cap (\ts \times \ts )$,

(ii) $\tred {S}\cap (\ts \times \ts )\subset  \tred {R}\cap (\ts \times \ts )$,

(iii) $\tred {R}\cap (\ts \times \ts )=  \tred {S}\cap (\ts \times \ts )$,

(iv) $\tred {R}\cap (\ts \times \ts )\Join \tred {S}\cap (\ts \times \ts )$.

\end{cons}

One can easily show  the following result applying Theorem 
\ref{fotetel}, Observation \ref{pollak}, 
and 
Consequence \ref{gabor}.
\begin{cons}\label{zoltan}
Let $R$ be a left-linear GSM-TRS over $\S$.
Moreover,
let $g\in \S-\S_0$ such that $g$ does not occur on the left-hand side of 
any rule in $R$, and let $\sharp\in \S_0$ be irreducible for $R$.
Then 
it is decidable whether 
$R$ is left-to-right ground minimal.
\end{cons}

By Statement \ref{zoli} and Theorem \ref{fotetel} we have the following.
\begin{tet}
Each of the following questions is 
undecidable for any convergent  left-linear  
EPR-TRSs 
$R$ and $S$ over a
 ranked alphabet $\Omega$, for any recognizable tree language 
$L\subseteq T_\Omega$
given by a tree automaton over $\Omega$ recognizing $L$,
 where $\Gamma\subseteq \Omega$ is the smallest ranked alphabet for which
$R(L)\subseteq T_\Gamma$.

(i) Is $R(L)\cap S(L)$ empty?

(ii) Is $R(L)\cap S(L)$ infinite?

(iii) Is $R(L)\cap S(L)$ recognizable?

(iv) Is $T_\Gamma - R(L)$ empty?

(v) Is $T_\Gamma - R(L)$  infinite?

(vi) Is $T_\Gamma - R(L)$  recognizable?

(vii) Is $R(L)$ recognizable?

(viii) Is $R(L)=S(L)$?

(ix) Is  $R(L)\subseteq S(L)$?

\end{tet}


\begin{lem}\label{zsomle}
Let $R$ and $S$ be linear collapse-free rewrite systems over the disjoint ranked alphabets 
$\S$ and $\D$, respectively. 
Let $\Gamma$ be a ranked alphabet with $\S\cup\Delta\subseteq \Gamma$.
Consider $R$ and $S$ as rewrite systems over $\Gamma$.
Then 

(i) $\red S \c \red R\subseteq  \red R \cup 
(\red R \c \red S)$, and

(ii) $\tred {R\cup S}= \tred R \c \tred S$.
\end{lem}
\pr The proof of (i) is straightforward.
Condition (ii) is a simple consequence of (i).
\hfill $\b$

\begin{tet}\label{vuk}
Let $R$ be a linear collapse-free EPRF-TRS
and   $S$ be a linear collapse-free EPR-TRS
 over the disjoint ranked alphabets 
$sign(R)$ and $sign(S)$, respectively. 
Then $R\oplus S$ is a   linear collapse-free  EPR-TRS. 
\end{tet}
\pr Apparently, $R\oplus S$ is a   linear collapse-free  TRS. 
Let $L$ be a recognizable tree language over some ranked alphabet 
$\Gamma$, where $sign(R)\cup sign(S)\subseteq \Gamma$.
By Lemma \ref{zsomle}, 
$(R\oplus S)^*_\Gamma(L)= S^*_\Gamma(R^*_\Gamma(L))$.
As $R$ is an EPRF-TRS,  $R^*_\Gamma(L)$ 
is recognizable. Moreover, since $S$
preserves recognizability, $S^*_\Gamma(R^*_\Gamma(L))$ is also recognizable. 
\hfill $\b$

\noindent
The proof of the following result is similar to the proof of Theorem
\ref{vuk}.
\begin{tet}
Let $R$ be a linear collapse-free PRF-TRS
and   $S$ be a linear collapse-free PR-TRS
 over the disjoint ranked alphabets 
$\S$ and $\D$, respectively. 
Then $R\oplus S$ is a PR-TRS. 
\end{tet}

\begin{tet}\label{zuk}
Let $R$ and $S$ be TRSs over the disjoint ranked alphabets 
$\S$ and $\D$, respectively,  such that any left-hand side in $R\oplus S$
differs  from a variable.
If $R\oplus S$  is an EPRF-TRS, then  $R$ and $S$ are also EPRF-TRSs. 
\end{tet} 
\pr Let $L$ be a finite recognizable tree language over some ranked alphabet $\Gamma$,
where $\S\subseteq \Gamma$. It is sufficient to show that 
$R^*_\Gamma(L)$ is recognizable.
Without loss of generality we may   rename the symbols of $\Gamma$
such that $\Gamma\cap \D=\emptyset$. 
Thus $R^*_\Gamma(L)=(R\oplus S)^*_{\Gamma\cup \Delta}(L)$. 
Since $\S\cup \Delta \subseteq \Gamma\cup \Delta$ and
 $R\oplus S$ is an EPRF-TRS, we get that 
$R^*_\Gamma(L)$ is recognizable and we can effectively construct a bta recognizing $R^*_\Gamma(L)$.
\hfill $\b$

\section{Conclusion and Open Problems}\label{geza}
We showed that each left-linear GSM-TRS is an EPRF-TRS.
We showed that  reachability, joinability, and local confluence are    decidable for 
  EPRF-TRSs. 
 We showed that the following problem is undecidable:

{\em  Instance:} A  
 murg TRS $R$   over  a ranked alphabet $\S$.

{\em Question:} Is  $R$ a P$\S$RF-TRS?

\noindent
Our results give rise to several open problems.

$\bullet$ Generalize the notion  of a left-linear  GSM-TRS such that the obtained TRS is still an EPRF-TRS.

$\bullet$ Show the following conjecture.
 Let $R$ be a right-linear TRS  over $sign(R)$,  and let $\S=\{\,g,\sharp\,\}\cup
sign(R),$ where $g\in\S_1-sign(R)$ and $\sharp\in\S_0-sign(R).$ Then $R$ is an 
 EP$\S$RF-TRS  if and only if $R$ is an EPRF-TRS. 
Show the corresponding conjectures when $R$ is  left-linear or  $R$ is  linear.

$\bullet$ Show that 
a string rewrite system $R$ over the alphabet  $alph(R)$ of $R$ preserves $alph(R)$-recognizability of finite string languages if
 and only if $R$ preserves recognizability of finite string languages.
Show that it is not decidable for a   string rewrite system $R$ whether $R$ preserves  $alph(R)$-recognizability of finite string languages,
 and 
whether $R$   preserves recognizability of finite string languages. 
Hence it is not decidable for a linear TRS $R$ whether $R$ is a  P$\S$RF-TRS and  whether $R$ is a  PRF-TRS.

$\bullet$ Show that
the property preserving recognizability of finite tree  languages  and the property effectively  
preserving recognizability of finite tree  languages
are modular   for the class of all left-linear collapse-free TRSs, for the class of all right-linear collapse-free TRSs, 
for the  class of all linear collapse-free TRSs, and for the  class of all  collapse-free TRSs.

\end{document}